\documentstyle[12pt,psfig]{article}
\newcommand{\eq}[1]{(\ref{#1})}
\newcommand{\nn}{\nonumber}
\newcommand{\fr}{\frac}
\newcommand{\gp}{p\!\!\!/}

\begin{document}
\topmargin 0pt
\oddsidemargin 1mm
\begin{titlepage}
\begin{flushright}
 KEK Preprint 96-129\\
 KEK-TH-496 \\
 OU-HET 251\\
 22 Nov. 1996 
\end{flushright}

\setcounter{page}{0}
\vspace{2 mm}
\begin{center}
{\Large Thermodynamic properties of spontaneous magnetization 
               in Chern-Simons $QED_3$}
\end{center} 
\vspace{6 mm}
\begin{center}
{\large  Shinya Kanemura 
  \footnote{e-mail: kanemu@theory.kek.jp}}\\
\vspace{2mm}
{\em Theory Group, KEK,\\
       Tsukuba, Ibaraki 305, Japan}\\
and  \\
\end{center}

\begin{center}
{\large Takao Matsushita
\footnote{e-mail: tmatsu@funpth.phys.sci.osaka-u.ac.jp}}\\
\vspace{2mm}
{\em Department of Physics, Osaka University,\\
              Toyonaka, Osaka 560, Japan}\\
\end{center}

\vspace{4 mm}

\begin{abstract}
The spontaneous magnetization in Chern-Simons $QED_3$ 
is discussed in a finite temperature system. 
The thermodynamical potential is analyzed within the weak field approximation 
and in the fermion massless limit. 
We find that there is a linear term with respect to the magnetic field 
with a negative coefficient at any finite temperature. 
This implies that the spontaneous magnetic field does not 
vanish even at high temperature.
In addition, we examine the photon spectrum in the system. 
We find that the bare Chern-Simons coefficient is 
cancelled by the radiative effects.
The photons then become topologically massless  
according to the magnetization, though they are massive by   
finite temperature effects.  
Thus the magnetic field is a long-range force without the screening 
even at high temperature.
\end{abstract}

\vspace{22mm}

\end{titlepage}

\section{Introduction}

\hspace*{18pt}
The 2+1 dimensional gauge field theories with Chern-Simons (C-S) term 
have had a lot of motivation to study. 
In addition to their own theoretical interests, they are expected 
to describe the physics of the planar systems like the quantum 
Hall effect \cite{qh1,qh2} and the high $T_C$ superconductivity \cite{htc}. 
Many interesting and remarkable properties of these theories 
have been found in past a decade, such as their chiral dynamics \cite{ch1,ch2}, 
the spontaneous parity violation by radiatively induced C-S terms \cite{pb}.    
A few years ago, it was also found that in a model of 
three dimensional quantum electrodynamics ($QED_3$) with a 
bare C-S term the spontaneous magnetic field can be stable 
by the fermion loop effects, which eventually triggers the 
dynamical Lorentz symmetry breaking (Hosotani \cite{Ho}).  
He has calculated the energy densities as a function of 
the magnetic field $B$ in the fermion one-loop level and 
showed that the minimum of the energy density can be  
at $B \neq 0$ under a consistency condition for non-zero magnetic fields.     
It has been found that this condition is also regarded as  
the sufficient condition for the spontaneous generation of the magnetic 
field  as well \cite{Ho2}.
According to the spontaneous Lorentz symmetry breaking, the physical photon,  
which is topologically massive in the tree level due to the bare C-S term, 
becomes massless as a Nambu-Goldstone boson.  
This mechanism of the spontaneous magnetization 
has been extended into the cases with massive fermions with  
non-zero fermion density \cite{IS}. 
Even in the case of no bare C-S term, the spontaneous magnetization 
can occur by adding heavy fermions 
because  the C-S term induced by them 
can play the same role as the bare C-S term \cite{Ho3}.   
The spontaneous magnetic field is expected to be important in the connection 
with the problem of the fractional quantum Hall effect \cite{qh2}.

In this paper, the thermodynamic properties of the spontaneous magnetization 
in $QED_3$ with a bare C-S term are investigated.
Some of the thermodynamic properties of the model 
without external fields have been 
studied to some extent in the different contexts so far.       
The specific heat for different values of the statistical angle 
is analyzed in ref. \cite{br}. 
The restoration of the parity at high temperature limit  
has been discussed in ref. \cite{KK}, in which it has been found that 
the radiatively induced parity violating C-S term 
vanishes only in the high temperature limit. 
Our aim here is to examine whether the spontaneous magnetic field, 
which is dynamically generated at $T=0$, 
vanishes or not in a high temperature region. 
The model we are studying here is based on ref. \cite{Ho}, 
in which all fermions are very light so that a non-zero bare C-S term 
is needed in the Lagrangian for the magnetization \cite{Ho3}.   
In the finite temperature system, the energy density 
is replaced by the thermodynamical potential.
The quantum fluctuation of the thermodynamical potential 
is investigated in the fermion one-loop level here. 
The coefficient of the linear term with respect to the magnetic field 
in the potential is calculated as a function of temperature  
\footnote{In our previous preprint \cite{km1}, some insufficient treatment 
has been done especially in the evaluation of the 
contribution of the zero-mode to the coefficient. 
These are correctively changed in the first half of this paper.} 
as well as the classical part of the potential.
Negative values of the coefficient mean the existence 
of the spontaneous magnetic field at finite temperature.
We find that the coefficient becomes more (negative) large 
behaving like $\sim - \log T$ in  high temperature region.    
Thus the spontaneous magnetic field does not vanish even at high temperature.

In addition,  the spectrum of photons  
is also investigated in the finite temperature system. 
At zero temperature, the photons are topologically massive at tree level. 
The mass amounts to the C-S coefficient.   
If the magnetization occurs,  
the bare coefficient of the C-S term is exactly cancelled by 
the radiative effects \cite{cht,bal}, so that 
the photons becomes massless. 
Therefore the photons are regarded as 
the Nambu-Goldstone bosons of the dynamical breakdown of the Lorentz symmetry 
\cite{Ho,Ho2}.
On the other hand, at finite temperature, 
the Lorentz symmetry is explicitly broken by thermal effects. 
The photons are then no longer massless even if the magnetization occurs. 
In general, they eventually have the thermal masses \cite{lin,ft1,wel}. 
The magnetic part of the field strength, however,  become statically massless 
and the magnetic force is a long-range force without the screening 
\cite{KK} if the topological mass is cancelled out. 
We find that, in our model with massless fermions, 
the bare topological mass is completely cancelled by the radiative effects 
even at finite temperature if the magnetization occurs.
This implies that the static mass of magnetic fields vanishes according 
to the magnetization even at high temperature. 
This is consistent with the result of the non-vanishing magnetic 
field by the thermodynamic potential approach.
It is also found that in the case of massive fermions 
the topological mass is cancelled out    
only in the `chirally symmetric' cases with massive fermions.
 
Furthermore we naively realize the reason of the spontaneous magnetization 
from the viewpoint of the difference of the free energies of 
free photons between $B \neq 0$ and $B = 0$, where 
$B$ is the magnetic field.    
It is found from rough estimation that 
the linear $|B|$ term with a negative coefficient appears 
in the difference. This explains the magnetization at finite temperature 
to some extent.  
These characteristics of the magnetization at finite temperature 
might be interesting 
in the connection with the physics such as the high $T_C$ superconductivity.

This paper is organized as follows.
In the next section, some general properties and 
statistic characteristics of the model will be  presented. 
In Sec 3, the spontaneous magnetization in this model in $T=0$ case 
will be reviewed briefly for a preparation for $T \neq 0$ cases 
in order to clarify the essential point for the magnetization 
and also to fix some notations.  
The strategy to extend the investigation to the finite temperature 
system will be explained then. 
Sec. 4  will be devoted to the calculation of self-energies     
of gauge bosons in the one-loop level and the weak-field approximation. 
In Sec. 5, the results that the spontaneous magnetic field does not vanish 
will be presented in the fermion massless limit 
by analyzing the effective potential both analytically and numerically.      
In Sec. 6, the spectrum  of the physical photons will be examined 
in the finite temperature system.
In Sec. 7, the reason why the spontaneous magnetic field does not vanish 
will be  discussed. Also some comments on our results will be in order.   
In the last section, we will summarize our results.  
Some definitions, details of derivations
and formulas are summarized in Appendices.

\section{Model in Finite Temperature System}

\hspace*{18pt}
We consider the model of $QED_3$ with a bare Chern-Simons term 
described by the Lagrangian 
\begin{eqnarray}
{\cal L} = - \frac{1}{4} F^{\mu \nu} F_{\mu \nu}
           - \frac{\kappa}{2} \epsilon^{\mu \nu \rho} 
             A_{\mu}\partial_{\nu}A_{\rho}
           + \sum_{a}\bar{\psi}_a 
          \left\{
             \gamma_a^{\mu}(i\partial_{\mu}+q_a A_{\mu})- m_a
          \right\}
             \psi_a  \,,    \label{lag}
\end{eqnarray}
where gamma matrices are defined as 
$\gamma_a^{\mu} \equiv (\eta_a \sigma^3 , i \sigma^1 ,i \sigma^2 )$. 
There can be two types of photons according to the choice of  
$\eta_a \equiv (i/2){\rm tr}\gamma_a^0\gamma_a^1\gamma_a^2 = \pm 1$.
Since the model \eq{lag} has charge conjugation invariance, 
we can take the electric charges $q_a$ to be positive without loss of generality.
Also since the transformation $m_a \leftrightarrow - m_a$ is equivalent to 
$\eta_a \leftrightarrow - \eta_a$, we can consider $m_a$ as non-negative. 
We then call the fermion with $\eta_a = \pm 1$ as $\eta_{\pm}$-fermion, 
respectively.

Under the existence of a classical magnetic field $B$  
($\langle 0 | A^{\mu} |0 \rangle  = - \delta^{\mu 1} x^2 B$),  
the energy spectrum for the $\eta_+$ fermion takes the form \cite{lag}
\begin{eqnarray}
  \begin{array}{l}
     E_0^a =  \epsilon (q_a B) \times \omega^a_0,    \\
     E_n^a = \pm \; \omega_n^a, \;\; (n \geq 1),\\ 
  \end{array} 
\end{eqnarray}
where $\epsilon (x)$ is the usual sign function, 
$\omega_n^a = \sqrt{m^{2}_a + 2 n / l_a^2}$ and $l^2_a = 1/|q_a B|$.
Then the general solution of the Dirac equation is 
\begin{eqnarray}
\psi_a (x) &=& \sum_{n,p}
             a_{np} 
             \left\{ \begin{array}{c} u_{np}^a(x)   \\
                                      w^{ac}_{np}(x)  \end{array}  \right\} +    
             \sum_{n,p} 
             b_{np}^{\dagger} 
             \left\{ \begin{array}{c} w^a_{np}(x)   \\
                                      u^{ac}_{np}(x)  \end{array}  \right\},  
        \left. \begin{array}{c}   (q_a B  > 0)   \\
                                   (q_a B  < 0) 
         \end{array} \right.,  \label{fer}
\end{eqnarray}
where $u^c_{np} = U_c \bar{u}_{np}^t, U_c = \gamma_2$ and the same 
for $w^c_{np}$. 
The positive energy solution $u_{np}$ and the negative energy solution $w_{np}$ 
in eq. \eq{fer} are  
\begin{eqnarray}
  u_{0p}(x) &=& \fr{1}{\sqrt{lL_1}} e^{-i(\omega_0 t + k x^1)} 
        \left(   \begin{array}{c}   v_0(\xi) \\
                                0 
                                   \end{array}\right),   \label{ux0} \\
  u_{np}(x) &=& \fr{1}{\sqrt{lL_1}}  e^{-i(\omega_n t + k x^1)} 
        \fr{1}{\sqrt{2\omega_n}}  
        \left(   \begin{array}{c}   \sqrt{\omega_n + m} v_n(\xi) \\
                                -i \sqrt{\omega_n - m} v_{n-1} (\xi)
                                   \end{array}\right),   (n \geq 1 ),
\label{uxn} \\
  w_{np} (x) &=& \fr{1}{\sqrt{lL_1}} e^{+ i( \omega_n t - k x^1)}
      \fr{1}{\sqrt{2\omega_n}}  
        \left(   \begin{array}{c}   \sqrt{\omega_n - m} v_n(\xi) \\
                                i \sqrt{\omega_n + m}  v_{n-1} (\xi)
                                   \end{array}\right), \;  
(n \geq 1), \label{swx}
\end{eqnarray}
where we dropped the fermion index $a$ for simplicity, and 
$k = 2 \pi p / L_1$ ($p$: integer), $\xi = x^2 / l - l k$ and 
$v_n(\xi) = (-1)^n \pi^{-1/4}(2^n n!)^{-1/2} 
e^{\xi^2/2} d^n/d \xi^n e^{-\xi^2}$. 
The solution for $\eta_-$ fermion is obtained by making   
the transformation $(t \rightarrow - t)$ from corresponding $\eta_+$ cases.
There is an asymmetry between positive and negative energy solutions 
in the lowest Landau levels. 
The vacuum expectation value of the total charge is (see eq. \eq{cha})  
\begin{eqnarray}
\langle 0 | \hat{Q} | 0 \rangle
= \sum_a q_a N_s^a \eta_a \epsilon (B)
            \left( \nu_a - \frac{1}{2} \right),        \label{chv}
\end{eqnarray}
where 
the number of degeneracy is given by 
$N_s^a = \sum_p 1 =  |q_a B|/2 \pi \times \int d^2x$ and  
we introduced the filling factor 
$\nu_a \equiv \langle 0 | \hat{\nu}_a | 0 \rangle$ for each fermion $\psi_a$,
\begin{eqnarray}
   \hat{\nu}_a = \left\{   \begin{array}{c}
   \sum_p  a^{\dagger}_{0p}a_{0p}  / N_s^a, 
   \;\;\; ( \eta_a \epsilon (B) > 0),                               \\ 
   \sum_p  b^{\dagger}_{0p}b_{0p} / N_s^a, 
    \;\;\; ( \eta_a \epsilon (B) < 0). 
                     \end{array}        \right.
\label{fil}
\end{eqnarray}
We here consider the completely filled ($\nu_a = 1$) 
or the empty ($\nu_a =0$) cases for zero temperature.    
From the time component of the equation of motion with the charge, 
we have a relation $\int d^2x\kappa B = \langle 0 | \hat{Q} | 0 \rangle$. 
The consistency condition for $B \neq 0$ is then obtained from this
relation and eq. \eq{chv} as 
\begin{eqnarray}
\kappa = \sum_a \frac{\eta_a q_a^2}{2\pi} 
       \left( \nu_a - \frac{1}{2} \right).
 \label{cc2}
\end{eqnarray}

In this paper, we would like to study this model 
in the finite temperature system.  
The partition function for the fermion sector is given by 
\begin{eqnarray}
  Z &=& {\rm tr} e^{- \beta( \hat{H} - \mu \hat{Q} )} \nonumber \\
    &=& \prod_a \left( e^{ - \beta \omega_0^a} 
                       e^{\frac{1}{2}\beta \mu_a q_a \eta_a \epsilon (B)} +  
          e^{-\frac{1}{2}\beta \mu_a q_a \eta_a \epsilon (B)}\right)^{N_s^a}
        \nonumber \\
    & & \times 
        \prod_{n=1}^{\infty}
        \left( 1 + e^{-\beta (\omega_n^a - \mu_a q_a)} \right)^{N_s^a}
        \left( 1 + e^{-\beta (\omega_n^a + \mu_a q_a)} \right)^{N_s^a}
        e^{\beta \omega_n^a N_s^a} ,
\label{par}
\end{eqnarray}
where $\beta = 1/ T$ and $\mu \hat{Q} = \sum_a \mu_a \hat{Q}_a$ 
(see eqs. \eq{ham} and \eq{cha}).
The total charge $Q = \sum_a Q_a$ 
in the finite temperature system is obtained by using \eq{par},
\begin{eqnarray}
 Q_a    = \frac{1}{\beta}
         \frac{\partial \ln Z}{\partial \mu_a}. \label{dmu}
\end{eqnarray}
Note that our definition of the chemical potential is a little  
different from the ordinary one and $q_a \mu_a$ in our notation 
is normally called the chemical potential in the literature. 
The condition for $B \neq 0$ in the finite temperature system 
is given from the time component of the equation of motion as 
$\kappa  = Q /(VB)$, where $V$ is the two-dimensional volume.  
In the massless limit ($m_a \rightarrow 0$), this becomes 
by using eqs. \eq{par} and \eq{dmu} 
\begin{eqnarray}
  \kappa &=& \epsilon (B) \sum_a \fr{q_a^2}{2 \pi}
  \left\{ 
     \frac{1}{2} \tanh \left( \frac{1}{2} q_a \beta \mu_a   \right)  \right.
 \nonumber         \\
 & & \;\;\;\;\;\;\;\;\;\left. + 
 \sum_{n=1}^{\infty} \left(  \frac{1}{e^{\beta (\omega^a_n - q_a \mu_a )}+1 }
  -   \frac{1}{e^{\beta (\omega^a_n + q_a \mu_a )}+1 } \right)
\right\}.
\label{id3}
\end{eqnarray}

The chemical potential $\mu_a$ is determined as a function of $T$ and $B$ 
by eq. \eq{id3} or 
by the following relations which are induced from eqs. \eq{cc2} and \eq{id3}:
\begin{eqnarray}
 1 &=&  \tanh \left( \frac{1}{2} q_a \beta |\mu_a| \right) 
\nonumber         \\
 & &  + 
 2 \sum_{n=1}^{\infty} \left( 
 \frac{1}{e^{\beta (\omega^a_n - q_a |\mu_a| )}+1 }  - 
  \frac{1}{e^{\beta (\omega^a_n + q_a |\mu_a| )}+1 } \right),
\label{id4}
\end{eqnarray} 
and $\epsilon (\mu_a) = 2 \epsilon (B) \eta_a (\nu_a - 1/2)$.
\begin{figure}[t]
\centering{
\leavevmode
\psfig{file=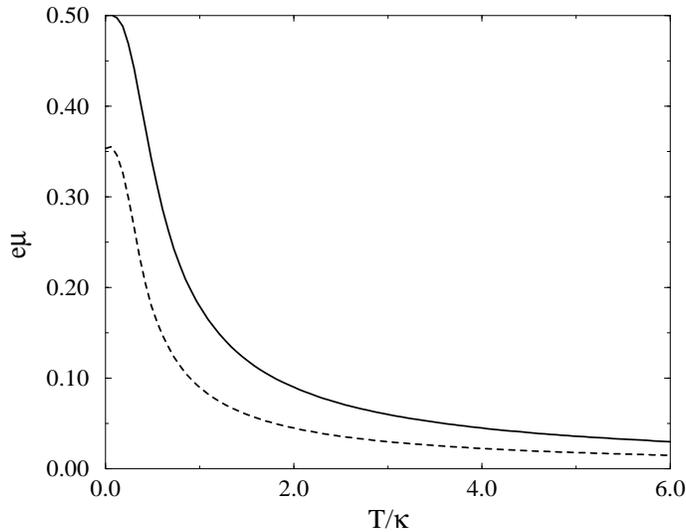,height=80mm,width=100mm,angle=-90}
\vskip -0mm }
\caption{The chemical potential $|\mu_a|$ in the fermion massless limit. 
         By assuming the consistency condition (12), $\mu$ become a function 
         of $B$ and $T$.
         It can be seen that $|\mu_a|$ is a monotonous decreasing 
         function of $T$ with the definite magnetic field $|B|$. 
         The solid line is  the case of $|q_a B|/{\kappa^2}=1/2$.
         The dashed line is the case of $|q_a B|/{\kappa^2}=1/4$. }
\end{figure}
The asymptotic behavior of $|\mu_a|$ is obtained from \eq{id4}.
We can easily see that $|\mu_a| \simeq \omega_1^a/2 = \sqrt{|q_a B|/(2 q_a^2)}$
in the low temperature limit ($\beta \rightarrow \infty$). 
At high temperature, we also see that $|\mu_a|$ behaves like $\sim \beta |B|$.
We show the behavior of the chemical potential as a function of $T$ at  
some values of $B$ (Figure 1). It can be seen in the figure that $|\mu_a|$  
is a monotonously decreasing (increasing) function of $T$ for fixed 
$B > 0$ ($B < 0$) and is also a increasing function of $|B|$ with fixed $T$. 
We also find from the relation \eq{id4} that for small $\beta^2 |B|$,
\begin{eqnarray}
  |\mu_a| \simeq \fr{1}{4 \ln 2} \beta |B| + {\cal O}(|B|^2).  \label{bzm}
\end{eqnarray}
Note that the non-zero $\mu_a$ here comes from the asymmetry of the 
lowest Landau levels. 
In the case of $B=0$, the asymmetry vanishes so that $\mu_a$ becomes zero.

\begin{figure}[t]
\centering{
\leavevmode
\psfig{file=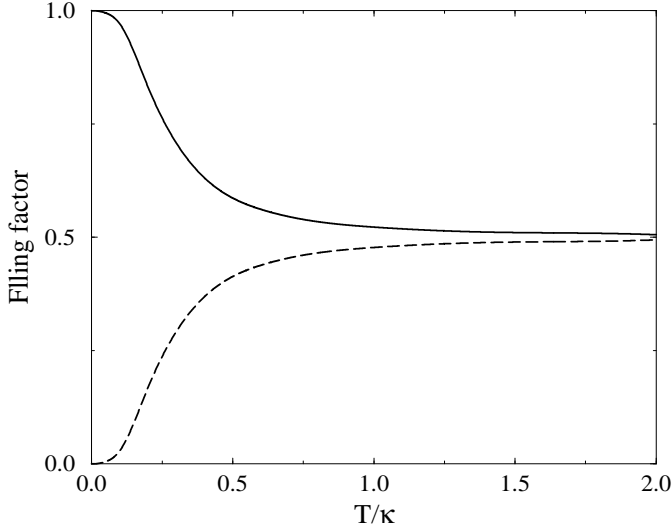,height=80mm,width=100mm,angle=-90}
\vskip -0mm }
\caption{The thermal behavior of the filling factors  
         $\langle \hat{\nu}^{\pm} \rangle$ for fixed $|B|$. 
         The solid line shows the behavior of $\langle \hat{\nu}^{+} \rangle $.
         The dashed line is that of $\langle \hat{\nu}^{-} \rangle $.
         We here set $\nu^+=1$ and $\nu^-=0$ at $T=0$, taking account of 
         eq. (9).}
\end{figure}
 Now we comment on the thermal behavior of the filling factors. 
The choice of the filling factors at zero temperature 
is taken to satisfy the consistency condition \eq{cc2} with $\kappa \neq 0$. 
The thermal average of filling factors is easily calculated by using \eq{fil} as
\begin{eqnarray}
  \langle \hat{\nu}_a \rangle = 
\frac{1}{e^{- \beta q_a \mu_a \eta_a \epsilon(B)} + 1}.
\label{fil2}
\end{eqnarray}
When we set $\nu^+$ and $\nu^-$ (the filling factors for $\eta_+$ and 
$\eta_-$-fermions, respectively) into 1 and 0 at $T=0$ respectively, 
the thermal average $\langle \hat{\nu}^+ \rangle$ 
($\langle \hat{\nu}^- \rangle$) becomes the monotonous decreasing 
(increasing) function of $T$ and 
$\langle \nu^{\pm} \rangle \rightarrow 1/2$ at large $T$ (see Figure 2).
This merely shows that the probability that the $\eta_{\pm}$-fermion exits 
in a lowest Landau level is multiplied by no  statistical weight 
at high temperature limit.

\section{Spontaneous Magnetization, Strategy to Finite Temperature System}

\hspace*{18pt}
In this section, we will show our strategy to study the spontaneous 
magnetization in the finite temperature system. 
To this aim, we start from a short review for the zero temperature case, 
which have been discussed at first by Hosotani \cite{Ho,Ho2}. 
Next we shall discuss the extension to the finite temperature case.   

At zero temperature, Hosotani has shown in the same model that  
the spontaneous magnetic field can be stable and then 
triggers the spontaneous breakdown of the Lorentz invariance. 
The effective potential has been calculated 
in a weak field approximation and the fermion massless limit \cite{Ho}.  
The classical part ($\Delta {\cal E}^{(3)}$ in ref. \cite{Ho}) 
of the potential yields the contributions of 
the orders of $|B|^{3/2}$ (matter) and $|B|^2$ (photon) 
with positive coefficients. 
The minimum of the potential of the classical part is then at $B = 0$. 
The radiative correction to the potential drastically changes the situation.
The deviation of the fermion-loop contribution to the effective potential 
between $B \neq 0$ and $B = 0$,  
${\cal E}^{\rm rad}$ ($\equiv \Delta {\cal E}^{(1)} - 
\Delta {\cal E}^{(2)}$, in ref. \cite{Ho}),  
is given  by  
\begin{eqnarray}
{\cal E}^{\rm rad} 
     = - \fr{i}{2} \int \fr{d^3p}{(2\pi)^3}
          \ln \fr{(1 + \Pi_0)
                  \left\{ 1 + (p_0^2 \Pi_0 - \vec{p}^2 \Pi_2)/p^2 \right\}
                    - (\kappa - \Pi_1)^2/p^2 }{(B \rightarrow 0)},\label{era}
\end{eqnarray}
where $\Pi_i ( = \sum_a \Pi_i^a), (i=0, 1, 2)$ 
are defined from the one-loop contribution 
of the vacuum polarization tensor:   
\begin{eqnarray}
\Gamma^{\mu\nu}(p) &=& (p^{\mu}p^{\nu} - p^2 g^{\mu\nu}) \Pi_0(p) + 
                       i \epsilon^{\mu\nu\rho} p_{\rho} \Pi_1(p) \nn \\
           & & + (1 - \delta^{\mu 0})(1 - \delta^{\nu 0})
                 (p^{\mu}p^{\nu} - \vec{p}^2 \delta^{\mu\nu})
                 (\Pi_2(p) - \Pi_0(p)).  \label{dpi}
\end{eqnarray}
In the case of $\forall q_a = e ( > 0)$ and $N_f^+ = N_f^- \equiv N$ 
(called the chirally symmetric case), putting the filling factors   
$\nu^+ = 1$ and $\nu^- = 0$ on taking account of 
the condition \eq{cc2}, ${\cal E}^{\rm rad}$ is then calculated 
in the fermion one-loop level by taking a weak field approximation 
and the fermion massless limit:       
${\cal E}^{\rm rad} = - e \kappa/\pi^2 \tan^{-1} (4/\pi ) \,|B| + 
\cdot\cdot\cdot$.
This implies that there cannot be the minimum value of the effective potential 
at the point of $B = 0$ so that the spontaneous magnetization occurs.
Then by the relation   
$\langle 0 | [ i M^{0i}, F^{0j}(x)] | 0 \rangle = 
              \epsilon^{ij} B \neq 0 $
where $M^{\mu\nu}$ are the generators of the Lorentz transformation, 
it is found that the non-vanishing magnetic field $B$ 
induces the spontaneous breakdown of the Lorentz invariance. 
Note that the essential point for this result is the existence of 
the non-vanishing negative coefficient of the linear $|B|$ term 
in the effective potential, which 
comes from the radiative correction part ${\cal E}^{\rm rad}$. 
If we set the bare C-S parameter $\kappa$ into zero, 
the linear term vanishes and the symmetry is restored.

From now, we discuss the extension of the effective potential 
to the finite temperature system.  
The effective potential (energy density), ${\cal E}(B)$, is  
then replaced by the thermodynamical potential 
$\Omega(B, T, \mu)$ 
\cite{lan}. In the case of $B =0$, 
Bralic {\it et. al.} has calculated 
the potential in this model in the different context \cite{br}. 
In the $B \neq 0$ case, the classical part of the fermion sector is given by 
\begin{eqnarray}
  \Omega_0^{\rm fermi} = - \fr{1}{V \beta} \ln Z, \label{fct}
\end{eqnarray} 
where $V$ is the two-dimensional volume and $Z$ is defined in eq. \eq{par}. 
We show an expression of $\Omega_0^{\rm fermi}$ in Appendix 1. 
We confirm that there is no linear $B$ term in $\Omega_0^{\rm fermi}$ there 
\cite{vol}.   
The quantum fluctuation part of $\Omega$ is more important here    
because we are interested in the spontaneous magnetic field for which 
the quantum part is essential in the case of $T=0$. 
The extension of this part to the finite temperature 
system is well performed by the Matsubara method \cite{mm}.
The fluctuation interaction part, $\Omega^{\rm rad}$, of the 
thermodynamical potential is then obtained from the eq. \eq{era} just  
by taking the replacement \cite{dj}
\begin{eqnarray}
  p^0  &\rightarrow& i p_3 = i \fr{2m\pi}{\beta}, \;(m: {\rm integer}), 
\label{mm1}\\
  \int  \fr{d p^0}{2\pi} &\rightarrow& \fr{i}{\beta} \sum_m.   \label{mm2}    
\end{eqnarray}
and $\Pi_i^a (p^2, B) \rightarrow 
\tilde{\Pi}^a_i(\vec{p}^2, B, \beta)_m,\;(i = 0, 1, 2)$ 
for bosonic loop integrations. 
As to the fermion loops included in $\tilde{\Pi}_i$, we employ 
the replacement with imposing anti-periodic boundary condition,
\begin{eqnarray}
 k^0  &\rightarrow& i k_3 = i \fr{(2n + 1)\pi}{\beta} + q_a \mu_a, 
                                  \;(n: {\rm integer}), \label{mm3}\\
  \int \fr{d k^0}{2\pi} &\rightarrow& \fr{i}{\beta} \sum_n.  \label{mm4}
\end{eqnarray}
Thus the deviation of the quantum fluctuation part of the 
potential is formally expressed  in the finite temperature system  by  
\begin{eqnarray}
\Omega^{\rm rad} &=&   \fr{1}{2\beta} \sum_m 
                        \int \fr{d^2 \vec{p}}{(2\pi)^2}  \nn \\
\times&\ln& \fr{  (1 + \tilde{\Pi}_0) 
   \left\{ 1 + (p_3^2 \tilde{\Pi}_0 
                     + \vec{p}^2 \tilde{\Pi}_2)/(p_3^2 + \vec{p}^{\;2}) \right\}
     + (\kappa - \tilde{\Pi}_1)^2 / (p_3^2 + \vec{p}^{\;2}) }
                     {(B \rightarrow 0)}. \nn \\
\label{ftp}
\end{eqnarray}

We, however, will not treat this expression directly. 
Since the non-zero coefficient of the linear $B$ term 
is essential for the spontaneous magnetization, 
what we have only to know in detail for our purpose   
is just  
the coefficient $C(\beta, \mu_0)$ in ${\Omega}^{\rm rad}$, 
\begin{eqnarray}
{\Omega}^{\rm rad} = C(\beta, \mu_0) |B| + 
({\rm higher}\;\; {\rm orders}\;\; {\rm of}\;\; |B|), 
\label{lin}
\end{eqnarray}
where $\mu_0 \equiv \mu(\beta)|_{B=0}$ is zero in the 
fermion massless limit because of eq. \eq{bzm}.   
We will concentrate our attention into 
$C(\beta) \equiv C(\beta, 0)$ from now.  
The coefficient is decomposed into two parts as
\begin{eqnarray}
C (\beta) &=& 
\left. \fr{\partial{\Omega}^{\rm rad}}{\partial B} 
\right|_{B=0, \mu=0}
+ 
\left. \fr{\partial{\Omega}^{\rm rad}}{\partial \mu_a} 
       \fr{\partial \mu_a} {\partial B} 
\right|_{B=0, \mu=0} \nn \\
  &\equiv& 
C_0(\beta) + C_{\mu}(\beta).  \label{coe1}
\end{eqnarray}
The first term is obtained from eq. \eq{ftp} as 
\begin{eqnarray}
C_0(\beta)       &=&
 \frac{1}{2 \beta}\sum_{m=-\infty}^{\infty}\int 
                       \frac{d^2\vec{p}}{(2\pi)^2}  
\left[  \frac{ \partial \tilde{\Pi}_0}{\partial B}
\left\{p_3^2 + \vec{p}^{\;2} 
        +(p_3^2 \tilde{\Pi}_0 + \vec{p}^{\;2} \tilde{\Pi}_2) \right\} 
\right.  \nn \\
&& \left. + (1 + \tilde{\Pi}_0)\left\{ p_3^2  
      \frac{ \partial \tilde{\Pi}_0}{\partial B} + 
             \vec{p}^{\;2}
      \frac{ \partial \tilde{\Pi}_2}{\partial B} \right\} 
- 2 (\kappa - \tilde{\Pi}_1) 
      \frac{ \partial \tilde{\Pi}_1}{\partial B}  \right]      \nonumber \\  
 & &  \;\;\;\; \times
       \left[ \left(1+\tilde{\Pi}_0
                       \right)
                 \left\{ (p_3^2+\vec{p}^2) + \right. \right. \nonumber \\
& &   \;\;\;\;\;\;\;\;
      \left. \left.   
                      \left.   
              p_3^2 \tilde{\Pi}_0    +
              \vec{p}^2 \tilde{\Pi}_2
\right\} + (\kappa - \tilde{\Pi}_1)^2 
\right]^{-1} \right|_{B=0,\mu=0}.  \label{coe2}  
\end{eqnarray}
The second term in eq. \eq{coe1} 
is  the contribution of the chemical potential $\mu$.
We however easily see that  $C_{\mu}(\beta)$ does not contribute. 
Since $\Omega^{\rm rad}$ is the deviation of the quantum fluctuation between 
at $B \neq 0$ and at $B=0$, we can easily see that 
$C_{\mu}(\beta)$ becomes zero by virtue of eq. \eq{id4}.

As long as the coefficient $C(\beta)$ is negative, 
the spontaneous magnetic field does not vanish.
In order to evaluate the behavior of $C(\beta)$ by using \eq{coe2}, 
we need to calculate $\tilde{\Pi}_i$'s.  
In the next section, we will show the details of the 
calculation of $\tilde{\Pi}_i$'s. 
Since we are interested in the coefficient of the linear $|B|$ term, 
 $C_0(\beta)$, the calculation will be performed up to the order 
${\cal O}(B^1)$ there.

\section{Boson Self-energies at finite temperature}

\hspace*{18pt}
This section will be devoted to  the  calculation 
of the boson self-energies $\tilde{\Pi}_i (i = 0, 1, 2)$ at $T \neq 0$, 
which are composed of the fermion one-loop diagrams. 
In the $T=0$ field theory, 
the fermion propagator $S (x,y)$ in the classical field $B$
is  expressed by the proper-time method \cite{Ho}.
We just quote the expressions and some character of $S(x,y)$ in Appendix 2.     
The fermion one-loop contributions to the 
vacuum polarization tensors are then calculated by  
\begin{eqnarray}
\Gamma^{a\mu \nu}(p) 
           = i\,q^2_a \int \frac{d^3k}{(2\pi)^3}
                               {\rm tr} \left[ \gamma^{\mu}S_0^a(k)
                                               \gamma^{\nu}S_0^a
                                               (k-p)
\right],  \label{deg}
\end{eqnarray}
where propagator $S_0^a(p)$ is defined in eq. \eq{sft} and  
$\Gamma^{\mu \nu} ( = \sum_a \Gamma^{a \mu \nu})$  are related to 
$\Pi_i ( = \sum_a \Pi_i^a)$  in eq. \eq{dpi}.

Since we are interested in the coefficient of $|B|$ in $\Omega^{\rm rad}$, 
we here take the weak field approximation up to the order ${\cal O}(B^1)$ 
in the calculation with setting $\mu_a=0$. 
(In Sec. 6, we will adopt a little better approximation 
to evaluate the spectrum of the photons.) 
We here show it only in 
the case of $\eta^+1$-fermion in $q_a B > 0$ because 
other cases are connected with this case by eqs. \eq{s1} and \eq{s2}. 
At first, we consider the contributions of fermions with $\nu_a = 0$.  
We drop the fermion index $a$ for a while just for avoiding 
complexity.
The expansion of $S_0(p)_{\nu=0}^{\eta=+1}$ by $1/l^2 = |q B|$ becomes 
\begin{eqnarray}
S_0(p)_{\nu=0}^{\eta=+1}
&=&  \fr{m + \gp}{p^2 - m^2 + i \epsilon} 
   - \fr{m \sigma_3 + p_0 I }{ (p^2 - m^2 + i \epsilon )^2} 
     l^{-2} + {\cal O} (l^{-4})  \nn \\
&\equiv& S_0(p)^{\eta=+1(0)}_{\nu=0} 
   +  S_0(p)^{\eta=+1(1)}_{\nu=0} l^{-2} + {\cal O} 
     (l^{-4}). 
\label{exs}
\end{eqnarray}
We then obtain the expansion of $\Gamma^{\mu\nu}$ 
from eqs. \eq{deg} and \eq{exs} in the case of $\nu=0$: 
$\Gamma^{\mu\nu}(p)^{\eta=+1}_{\nu=0}
= \Gamma^{\mu\nu}(p)^{\eta=+1(0)}_{\nu=0} +
  \Gamma^{\mu\nu}(p)^{\eta=+1(1)}_{\nu=0}
  l^{-2} + {\cal O}(l^{-4})$,  where
\begin{eqnarray}
\Gamma^{\mu \nu}(p)_{\nu=0}^{\eta=+1(0)} 
     &\equiv& i\,q^2 \int \frac{d^3k}{(2\pi)^3}
       {\rm tr} \left[ \gamma^{\mu}S_0(k)^{(0)}_{\nu=0}
       \gamma^{\nu}S_0(k-p)^{(0)}_{\nu=0}
\right], \label{eg1} \\
\Gamma^{\mu \nu}(p)_{\nu=0}^{\eta=+1(1)}
   &\equiv& i\,q^2 \int \frac{d^3k}{(2\pi)^3}
      \left\{  {\rm tr} \left[ \gamma^{\mu}S_0(k)^{(0)}_{\nu=0}
          \gamma^{\nu}S_0(k-p)^{(1)}_{\nu=0}\right]\right. \nn \\
&& \;\;\;\;\;\;\;\;\;\;\;\;\;\;\;\;\;\;\;\;
\left. + {\rm tr} \left[ \gamma^{\mu}S_0(k)^{(1)}_{\nu=0}
                  \gamma^{\nu}S_0(k-p)^{(0)}_{\nu=0}
\right] \right\}.  \label{eg2}
\end{eqnarray}
Since we can easily check 
$\Gamma^{\mu\nu}(p)^{\eta=+1(1)}_{\nu=0}$ to be zero due to  
the properties of the Dirac $\gamma$-matrices, 
we can regard $\Gamma^{\mu\nu}(p)^{\eta=+1}_{\nu=0}$ as 
$\Gamma^{\mu\nu}(p)^{\eta=+1(0)}_{\nu=0}$ in this approximation.
Next, let us consider the case of $\nu=1$. 
The propagator in this case is given in \eq{sdc}.
It is convenient to see the deviation 
$ \delta \Gamma^{\mu\nu}(p)^{\eta=+1} \equiv
\Gamma^{\mu\nu}(p)^{\eta=+1}_{\nu=1} -  
\Gamma^{\mu\nu}(p)^{\eta=+1}_{\nu=0}$, 
where 
\begin{eqnarray}
\delta \Gamma^{\mu \nu}(p)^{\eta=+1}
   &=& i\,q^2 \int \frac{d^3k}{(2\pi)^3}
      \left\{  {\rm tr} \left[ \gamma^{\mu}S_0(k)_{\nu=0}^{\eta=+1}
          \gamma^{\nu} f (k-p) \right]\right. \nn \\
&& \;\;\;\;\;\;\;\;\;\;\;\;\;\;\;\;\;\;\;\;
\left. + {\rm tr} \left[ \gamma^{\mu}f(k)
                         \gamma^{\nu}S_0(k-p)_{\nu=0}^{\eta=+1}
\right] \right\} \nn \\
&& + i\,q^2 \int \frac{d^3k}{(2\pi)^3}
        {\rm tr} \left[ \gamma^{\mu} f(k)
          \gamma^{\nu} f (k-p) \right].\label{ex4}
\end{eqnarray}
Since  $f(p)$ includes the part 
$e^{- \vec{p}^{\;2}l^2}$ (see Appendix 2), 
it cannot be taken the Taylor expansion 
by $1/l^2$ in a usual sense. 
We have to keep this part in our calculation 
whereas we drop terms with the part 
$l^{-2} \times e^{- \vec{p}^{\;2}l^2}$. 
The last term of r.h.s. in eq. \eq{ex4} is then dropped in this sense. 
Hence, in our approximation, $\tilde{\Pi}_i$'s are 
calculated by the following expressions:   
\begin{eqnarray}
\Gamma^{\mu\nu}(p)^{\eta=+1}_{\nu=0}
&=&  i\,q^2 \int \frac{d^3k}{(2\pi)^3}
       {\rm tr} \left[ \gamma^{\mu}S_0(k)^{(0)}_{\nu=0}
       \gamma^{\nu}S_0(k-p)^{(0)}_{\nu=0}
\right]
 + {\cal O} (l^{-4}) , \label{ex2}\\
8\Gamma^{\mu \nu}(p)^{\eta=+1}_{\nu=1}
   &=& \Gamma^{\mu\nu}(p)^{\eta=+1}_{\nu=0}
+i\,q^2 \int \frac{d^3k}{(2\pi)^3}
      \left\{  {\rm tr} \left[ \gamma^{\mu}S_0(k)_{\nu=0}^{\eta=+1(0)}
          \gamma^{\nu} f (k-p) \right]\right. \nn \\
&+& \!\!\!
\left.  {\rm tr} \left[ \gamma^{\mu}f(k)
   \gamma^{\nu}S_0(k-p)_{\nu=0}^{\eta=+1(0)}\right] \right\} +   {\cal O} 
      ( l^{-2} e^{-l^2 p^2}).
\label{ex5}
\end{eqnarray}

Before proceeding to the calculation of $\tilde{\Pi}_i$, 
we mention the consistency of our approximation scheme.
The self-energies at $T=0$ ($\Pi_i$'s) are calculated by using 
the expressions in eqs. \eq{ex2} and \eq{ex5} 
up to the order ${\cal O}(B^1)$. 
They are presented in eqs. \eq{n1} $\sim$ \eq{n6} in Appendix 3. 
We can see that these expressions 
are consistent with the calculation by Hosotani \cite{Ho}.  
His calculation of the self-energies by virtue of
the proper-time representation 
includes the ultra-violet divergence which has to be renormalized. 
On the other hand, 
we do not have any ultra-violet divergence in our calculation. 
This is because the divergence part appears only 
in the term higher order than ${\cal O}(|B|)$, 
which we are now neglecting. 

Now we go to the case in the finite temperature system.
By applying the replacements \eq{mm3} and  \eq{mm4} 
(and $p_0 \rightarrow i p_3$)
to $\Pi_i$'s calculated in eqs. \eq{n1} 
$\sim$ \eq{n6} with eq. \eq{def}, 
we obtain $\tilde{\Pi}_0$, $\tilde{\Pi}_1$ and $\tilde{\Pi}_2$ 
 in the weak field approximation, 
which are presented in Appendix 4.
From now we take the fermion massless limit ($m_a \rightarrow 0$). 
We also restore the fermion index $a$ in the expression.  
For the case of $\nu_a = 0$, $\tilde{\Pi}_i$'s become then  
\begin{eqnarray}
\tilde{\Pi}^a_{0,\nu=0}(p_3, \vec{p}\,^2, B, \beta)
&=& \fr{q_a^2}{16} \fr{1}{(\vec{p}\,^2+p_3^2)^{1/2}} +
    A_0^a(p_3, |\vec{p}|, \beta)   + {\cal O}(|B|^2),    
\label{pt0}  \\
\tilde{\Pi}^a_{1,\nu=0}(p_3, \vec{p}\,^2, B, \beta) 
&=& 0 + {\cal O}(|B|^2),    \label{pt1}   \\
\tilde{\Pi}^a_{2,\nu=0}(p_3, \vec{p}\,^2, B, \beta)  
&=&  - \fr{p_3^2}{\vec{p}\,^{2}}
     \tilde{\Pi}^a_{0,\nu=0}(p_3, \vec{p}\,^2, B, \beta) \nn \\
&& \!\!\!\!\!\!\!\!\!\!\!\!
   + \fr{q_a^2}{16 \vec{p}^{\;2}} (\vec{p}^2+p_3^2)^{1/2} 
    - \fr{1}{\vec{p}\,^{2}} A_2^a(p_3, |\vec{p}|,\beta) + 
    {\cal O}(|B|^2),
\label{pt2}
\end{eqnarray}
where $A^a_0$ and $A^a_2$ are contributions of finite temperature effects from 
the fermion loops and take the forms as 
\begin{eqnarray}
& & A_0^a(p_3, |\vec{p}|, \beta)
  = \frac{q_a^2}{2 \pi \vec{p}^2}
    \int_0^{\infty} d k 
    \frac{1}{e^{\beta k}+1} \nonumber \\
& &\;\;\;\;\;\; \times
\left\{ 1 -  \left[ 
\frac{\sqrt{(\vec{p}^2 + p_3^2 - 4k^2)^2 + 16k^2p_3^2}
      + \vec{p}\,^2 + p_3^2 - 4k^2}
     {2(\vec{p}\,^2+p_3^2) }
\right]^{\frac{1}{2}} \right\},  \label{da1}\\
&&A_2^a(p_3, |\vec{p}|, \beta) 
 =  \frac{p_3^2 q_a^2}{2 \pi \vec{p}\,^2}
    \int_0^{\infty} d k 
    \frac{1}{e^{\beta k}+1} \nonumber \\
&& \;\;\;\;\;\;\;\;\;\;
    + \fr{q_a^2}{4\pi}\fr{(\vec{p}\,^2 + p_3^2)^{1/2}}
                 {\vec{p}\,^2}
     \int_0^{\infty} d k 
    \frac{1}{e^{\beta k}+1} \nonumber \\
&& \;\;\;\;\;\;\;\;\;
      \times \frac{(*)} {\sqrt{ (\vec{p}\,^2 + p_3^2 - 4k^2)^2 
                                     + 16 k^2 p_3^2}},
\label{da2}
\end{eqnarray}
where
\begin{eqnarray}
(*) &\equiv& \sqrt{2}\,
\left\{ (4 k^2 - p_3^2)
        \left[\sqrt{(\vec{p}\,^2+p_3^2-4k^2)^2 + 16 k^2 p_3^2 
     }+ \vec{p}\,^2+p_3^2-4k^2 
                \right]^{1/2}  \right. \nonumber \\
& & \; \left. - 4 k (p_3^2)^{1/2} \left[ \sqrt{ (\vec{p}\,^2 + p_3^2 - 4 k^2)^2 
                                              +16 k^2 p_3^2 } -
                        (\vec{p}\,^2 + p_3^2 - 4 k^2) 
                        \right]^{1/2}    \right\} \nn.
\end{eqnarray}
We can see from eqs. \eq{da1} and \eq{da2} that  
$\tilde{\Pi}^a_{0}$ and $\tilde{\Pi}^a_{2}$ grow like $\sim T$
at high temperature for fixed $p_3$.
This behavior has already been pointed out in ref. \cite{KK}, where 
the calculation has been based on the real-time formalism.  
As to the case $\nu=1$, we obtain  
\begin{eqnarray}
\tilde{\Pi}^a_{0,\nu=1}(p_3, \vec{p}\,^2, B, \beta) &=&  
\tilde{\Pi}^a_{0,\nu=0}(p_3, \vec{p}\,^2, B, \beta), 
\label{pt4}\\
\tilde{\Pi}^a_{1,\nu=1}(p_3, \vec{p}\,^2, B, \beta) &=& 
\eta_a \fr{q_a^3}{\pi} 
\fr{1}{\vec{p}\,^2 + p_3^2} |B| + {\cal O}(e^{- \vec{p}\,^2/(q_a|B|)}),
\label{pt5} \\ 
\tilde{\Pi}^a_{2,\nu=1}(p_3, \vec{p}\,^2, B, \beta) &=&  
\tilde{\Pi}^a_{2,\nu=0}(p_3, \vec{p}\,^2, B, \beta).
\label{pt6}
\end{eqnarray}
Note that only $\tilde{\Pi}^a_{1,\nu=1}$ depends on the magnetic field 
in this approximation. 
We can neglect the term of the order 
${\cal O}(e^{-\vec{p}\,^2/(q_a|B|)})$ 
in eq. \eq{pt5} for the evaluation of $C(\beta)$ 
because this term is sufficiently smaller than the linear term 
for small $|B|$.

Therefore, we have calculated all the self-energies $\tilde{\Pi}_i$ 
 up to the order ${\cal O}(B^1)$ in the fermion massless limit.
In the next section, we will investigate the character of the 
effective potential ${\Omega}^{\rm rad}$ by analyzing 
the coefficient of the linear term $C(\beta)$ by using    
the results  in \eq{pt0} $\sim$ \eq{pt6}.

\section{Thermodynamical Potential}

\hspace*{18pt}
We here examine 
the effective potential in the finite temperature system.
The purpose is, as we mentioned before, 
to decide  whether the spontaneous magnetic 
field, that takes non-zero value at $T=0$, vanishes at high temperature 
or not.      
We shall analyze the coefficient of the linear $|B|$ term  
$C_0(\beta)$ in the potential $\Omega (B, \beta, \mu_a)$ 
by using $\tilde{\Pi}_i$'s calculated in the massless limit in the last section. 
In the evaluation, we set  $q_a = e > 0$, and assume $N_f^+ = N_f^- = N$. 
The filling factors $\nu_+$ and $\nu_-$ are set their value into 1 and 0 
at $T = 0$, respectively.    
The condition \eq{cc2}  becomes then 
$\kappa = e^2 N/(2\pi)$. 
By using eqs. \eq{pt0} $\sim$ \eq{pt6} as well as this condition, 
the coefficient $C_0(\beta)$ in eq. \eq{coe2}
is expressed as 
\begin{eqnarray}
 C_0(\beta)= -\frac{e \kappa}{2 \pi^2} \alpha \sum_{m=-\infty}^{\infty} 
     \int_0^{\infty} d x \frac{x^5}{F_m(x\,,\,\alpha)}, \label{cot}
\end{eqnarray}
where
$\alpha \equiv 2 \pi T/\kappa$ and $x \equiv |\vec{p}|/\kappa$.
 
The function $F_m(x\,,\,\alpha)$ in r.h.s. of eq. \eq{cot} is defined as  
\begin{eqnarray}
F_m(x\,,\,\alpha) 
&\equiv&  \left[ x^2 \left(x^2+(\alpha m)^2 \right) 
              +\frac{\pi}{4} x^2 \sqrt{x^2 + (\alpha m)^2} 
             + \tilde{A}_0(x,\,\alpha,\,m)  \right]    \nonumber    \\  
& & \;\times
     \left[x^2 \left( x^2 + (\alpha m)^2 \right) 
            + \frac{\pi}{4} x^2 \sqrt{x^2 + (\alpha m)^2} 
              - \tilde{A}_2(x,\,\alpha,\,m) \right] \nonumber \\
& &     + x^4 \left( x^2 + (\alpha m)^2 \right)\,,
\label{dfn}
\end{eqnarray}
where
\begin{eqnarray}
  \tilde{A}_0(x,\alpha, m) &\equiv& 
       \sum_a A^a_0 
   \left(\kappa \alpha m, \kappa x, 2 \pi /(\kappa \alpha)\right) x^2 
         \left( x^2 + (\alpha m)^2 \right), \label{at1} \\
  \tilde{A}_2(x,\alpha,\,m) &\equiv& 
       \sum_a A^a_2 
   \left(\kappa \alpha m, \kappa x, 2 \pi /(\kappa \alpha)\right) 
         \left(\fr{x}{\kappa}\right)^2.\label{at2}
\end{eqnarray}

Before the numerical analyses, let us see the asymptotic behavior 
of $C_0(\beta)$ described in \eq{cot}.
At first, we consider the low temperature limit. 
Since $\tilde{A}_0$ and $\tilde{A}_2$  have the factor 
$1/(e^{\beta k} + 1)$ (see eqs. \eq{da1} and \eq{da2}), 
these terms are rapidly dumped. 
Then $F_m(x\,,\,\alpha)$ is replaced by $x^4 \times f(x\,,\,\alpha m)$, where
\begin{eqnarray}
f(x\,,\,\alpha m) = \left(x^2+(\alpha m)^2 \right)
    \left\{ 1 + \left( 
      \fr{\pi}{4} + \sqrt{x^2+(\alpha m)^2} \right)^2 \right\}.
\label{fli}
\end{eqnarray}
The coefficient is then calculated by making use of the Euler-Mclaurin's 
mathematical formula (see Appendix 6.) as below, 
\begin{eqnarray}
C_0 (\beta) &\rightarrow& - \fr{e \kappa}{2 \pi^2} \alpha 
       \left\{ 2 \sum_{m=0}^{\infty} \int_0^{\infty} d x \fr{x}{f(x,\alpha m)} 
                     - \int_0^{\infty} d x \fr{x}{f(x,0)}\right\} \nn \\
&=& - \fr{e \kappa}{2 \pi^2} \alpha
      \left\{ \fr{2}{\alpha} \int_0^{\infty}d y \int_0^{\infty} d x 
              \fr{x}{f(x,y)} + 
            \lim_{m \rightarrow \infty} 
         \int_0^{\infty} d x \fr{x}{f(x,\alpha m)} + 
     {\cal O}(\alpha) \right\} \nn \\
&=& - \fr{e \kappa}{\pi^2} \int_0^{\infty} d r 
         \fr{1}{1+ \left( \fr{\pi}{4} + r \right)^2} 
  + {\cal O}(\alpha^2), \;\;\; (x = r \cos \theta, y= r \sin \theta)\nn \\ 
&=& - \left\{ \fr{e \kappa}{\pi^2} \tan^{-1}\fr{4}{\pi} + {\cal O}(T^2)
      \right\}, 
  \;\; \left( T \simeq 0 \right). \label{cto}
\end{eqnarray}
Thus we indeed reproduce the result in \cite{Ho}.
Secondly,
at high temperature region, the contribution of the zero mode ($m=0$)  
becomes dominant  
because 
that of the non-zero modes ($m \neq 0$) is relatively suppressed by $1/T$ as  
\begin{eqnarray}
  C_0(\beta)_{m \neq 0 {\rm modes}} 
&=& - \fr{e \kappa}{\pi^2} \alpha \sum_{m=1}^{\infty} \int_0^{\infty}
    d x \fr{x^5}{F_m(x,\alpha)} \nn \\
&\rightarrow& - \fr{\kappa e}{6 \alpha} \int_0^{1} d t \fr{\sinh t}{\cosh^3 t}, 
  \;\;\;(x = \alpha m \sinh t) \nn \\
&=& - \fr{\kappa^2 e}{24 \pi} \fr{1}{T}, \;\;\;  (T \rightarrow \infty). 
\label{htn}
\end{eqnarray}
\begin{figure}[t]
\centering{
\leavevmode
\psfig{file=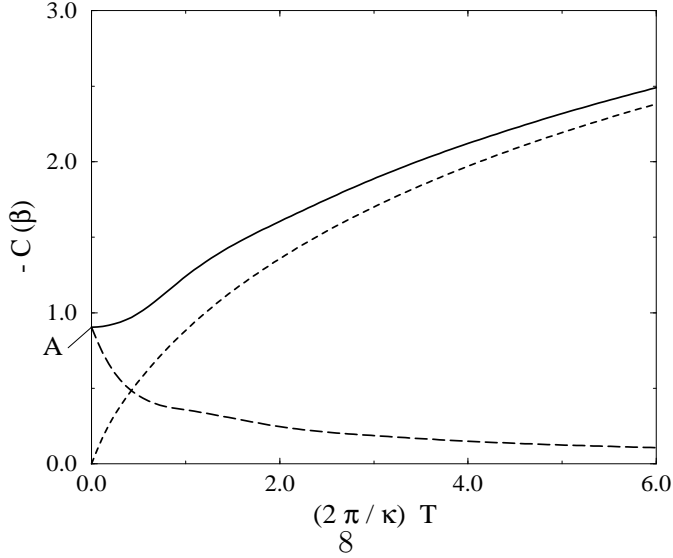,height=80mm,width=100mm,angle=-90}
\vskip -0mm}8
\caption{The behavior of the coefficient $C(\beta)$ 
         ($\beta = 1/T$) is shown by the solid line.
         It can be seen that the coefficient is negative and does not vanish 
         even at high temperature. Rather, the absolute value 
         becomes larger as  $\sim \log T$. 
         The point $A$ which is the low temperature limit
         takes the value $\kappa e/\pi^2 \tan^{-1}(4/\pi)$. 
         The dotted line is the contribution of  the zero-mode. 
         The dashed line is the summation of the other modes.}
\end{figure}
The contribution of the zero mode does not vanish in the limit and behaves like 
\begin{eqnarray}
  C_0(\beta)_{\rm m=0mode} \rightarrow  - 
\fr{e \kappa}{2\pi \ln 2} \ln \fr{T}{\kappa}, \;\;\; (T \rightarrow \infty). 
\label{htz}
\end{eqnarray}
Eqs. \eq{htn} and \eq{htz} show that $C_0(\beta)$ does not vanish at any finite 
temperature. 
The numerical analyses show that the coefficient $C_0(\beta)$ is always 
negative  at any temperature (Figure 3).  
The non-vanishing $C_0(\beta)$ means that 
the thermodynamical potential $\Omega$ does not have its minimum 
value at $B = 0$ at any temperature.

\section{Photon Spectrum at Finite $T$ and $\mu$}

\hspace*{18pt}
At $T=0$ and  $\mu = 0$ case the spontaneous magnetic field implies  
the dynamical Lorentz symmetry breaking.
The massless pole then appears in the spectrum of the physical photons  
by virtue of the Nambu-Goldstone theorem.  
Namely, there is no spontaneous magnetization at tree level and 
photons then have a topological mass $|\kappa|$  whereas 
the magnetization occurs at one loop level and photons then become
massless.  The massless-ness comes from the exact cancellation 
of the bare topological mass $\kappa$ by $\Pi_1(0)$ \cite{Ho,Ho2,bal}. 
It has been well known that the cancellation is exact for all orders of
the perturbation \cite{cht}. 

In the finite temperature system, the Lorentz symmetry is 
explicitly broken by the temperature effects.  
On the other hand, we have just shown by the thermodynamical potential
approach  that 
the spontaneous magnetic field does not vanish at any finite temperature.  
It is of very interest to examine the photon spectrum in the finite 
temperature system.  
The  photons then becomes massive by temperature effects even if 
the topological mass is cancelled out at loop levels.      
From the Lagrangian \eq{lag} the photon spectrum  
in the finite temperature system is determined by \cite{KK,wel} 
\begin{eqnarray}
  (p^2 - \tilde{\Pi}_L^2)(p^2 - \tilde{\Pi}_T^2) -
 p^2 \tilde{\Pi}_{\rm topo}^2 = 0,  \label{spe} 
\end{eqnarray}
where $\tilde{\Pi}_L$ and  $\tilde{\Pi}_T$ becomes the thermal masses for 
the longitudinal and transverse modes respectively when  the 
topological mass $\tilde{\Pi}_{\rm topo}$ is zero.
In our model $\tilde{\Pi}_{L,T}$ and $\tilde{\Pi}_{\rm topo}$ are 
\begin{eqnarray}
  \tilde{\Pi}_L(p^2) &=& - p^2  \sum_a \tilde{\Pi}^a_0(p^2), \\
  \tilde{\Pi}_T(p^2) &=& - p_0^2 \sum_a \tilde{\Pi}^a_0(p^2) + 
                                \vec{p}\,^2 
                                  \sum_a \tilde{\Pi}^a_2(p^2), \\
  \tilde{\Pi}_{\rm topo}(p^2) &=& \kappa - \sum_a \tilde{\Pi}_1^a(p^2),
\end{eqnarray}
where we wrote $p_3 = -i p_0$.
By using eqs. \eq{pt0}, \eq{pt2}, \eq{pt4} and \eq{pt6}, the 
expressions of $\tilde{\Pi}_L$ and $\tilde{\Pi}_T$ at high temperature 
are obtained in the chirally symmetric case as 
\begin{eqnarray}
  \tilde{\Pi}_L &=& 4 N \omega_p^2 \fr{p^2}{\vec{p}\,^2}
             \left(\fr{p_0}{p} - 1 \right),  \label{pl1}\\
  \tilde{\Pi}_T &=& 4 N \omega_p^2 \fr{p_0}{\vec{p}\,^2}
             \left( p_0 - p \right),  \label{pl2}
\end{eqnarray}
where 
$\omega_p^2 \equiv (e^2 \ln2/4\pi) T$ is the plasma frequency. 
These are consistent with the results by Klein-Kreisler {\it et. al.} 
\cite{KK}.

At first, let us consider the case of $\tilde{\Pi}_{\rm topo}(0) = 0$.
In terms of the field strength, $\tilde{\Pi}_L$  and $\tilde{\Pi}_T$ 
are the thermal masses of electric fields and magnetic fields
respectively.
In the static limit ($p_3 = 0, \vec{p} \rightarrow 0$) \cite{ft1}, 
 we easily obtain $\tilde{\Pi}_L = 4N \omega_p$ and  $\tilde{\Pi}_T = 0$. 
The electric fields are then massive and become short-range force 
and thus screened,  
whereas the magnetic fields are massless and remains long-range forces. 
In dynamic cases such as plasma oscillation, 
both fields are massive with the same mass $\sim \omega_p$.
Secondly, in the case of 
$\tilde{\Pi}_{\rm topo}(0) \equiv \tilde{\kappa} \neq 0$,      
the electric and magnetic fields become both massive. 
They have the same static masses $\sim \sqrt{4N \omega_p^2 + \tilde{\kappa^2}}$. 
Also, the dynamic masses of the photons become  approximately   
$\sim \sqrt{2N \omega_p^2 + \kappa^2}$.
Therefore if $\tilde{\kappa}$ is zero at any temperature, 
we can obtain a certain relationship at finite temperature 
between the magnetization and the statement that the magnetic fields  
remain long-range forces.

In the following we shall show that in our model, the topological mass 
$\tilde{\kappa}$ vanishes at any temperature: 
\begin{eqnarray}
  \tilde{\kappa} = \kappa - \sum_a \tilde{\Pi}^a_1(0) = 0. \label{kk}  
\end{eqnarray}
We stress that this is not contradictory with the results 
by Lykken {\it et. al.} \cite{lsw}. 
In our model, we at first put the fermion distribution at zero
temperature. The chemical potential is then not a free parameter but a function 
of $B$ and $T$, which is determined by the consistency condition at finite 
temperature.

Now we calculate $\tilde{\Pi}_1(0)$ in the weak field approximation.
we here leave the fermion massive (but not so heavy). 
For the $\nu=0$ fermions, the contribution is easily calculated 
from \eq{nu0} by putting $p_3, \vec{p} =0$ and using the formula \eq{def} as
\begin{eqnarray}
  \tilde{\Pi}_1^{a\nu=0}(0) = - \fr{q_a^2\eta_a}{4\pi}
                  \left( 1 - \fr{1}{e^{\beta (m_a - q_a\mu_a)} + 1} 
                           - \fr{1}{e^{\beta (m_a + q_a\mu_a)} + 1}  \right). 
\label{1n0}
\end{eqnarray}
We can easily see that, when $m_a =0$, $\tilde{\Pi}_1^{a\nu=0}(0)$ 
becomes zero at any temperature and density as in eq. \eq{pt1}. 
In the low temperature limit, eq. \eq{1n0} reproduces the result in 
\cite{ch2,Ho}.  
At high temperature limit, it becomes zero.    
Next, we  consider the case of $\nu=1$ fermions. 
We cannot use eq. \eq{pt5} to take the limit of $p_3, \vec{p} = 0$ 
because the approximation for eq. \eq{pt5} 
is of the order ${\cal O}(|B|)$ which is not appropriate here.  
So we go back to the expression in eq. \eq{n5}. 
Here we consider $\tilde{\Pi}_1^{a\nu=1}(0)$ as the form getting 
from applying the replacement 
\eq{mm1}, \eq{mm3} and \eq{mm4} to  eq. \eq{n5}.
By using the formula \eq{def}, we can decompose 
$\delta \tilde{\Pi}_1^{a} (\equiv 
\tilde{\Pi}_1^{a\nu =1} - \tilde{\Pi}_1^{a\nu =0})$ as 
\begin{eqnarray}
    \delta \tilde{\Pi}_1^{a}(p) =  \delta \Pi_1^{a}(p) + 
    \tilde{\Pi}^{a}_1(p, \beta)_{\rm Thermo}, 
\label{dc}
\end{eqnarray}
where  $\delta \Pi_1^a$ is the deviation 
between $\nu=0$ and $\nu=1$ of the induced C-S term at $T=0$, 
which has been calculated by using the proper time representation as \cite{Ho}, 
\begin{eqnarray}
   \delta \Pi_1^a (p) = \fr{\eta_aq_a^2}{2\pi} 
 \left( \fr{1}{2 - p^2 l_a^2 - 2m_a p_0 l_a^2} + 
        \fr{1}{2 - p^2 l_a^2 + 2m_a p_0 l_a^2}  \right). \label{dv}
\end{eqnarray}
In eq. \eq{dv}, when we neglect  the contribution of higher orders of $|B|$ 
than the first order, the result in eq. \eq{qq1} 
is of course reproduced.     
This becomes $\eta_a q_a^2/(2\pi)$ for $p_{\mu}=0$.
The second term in r.h.s. in eq. \eq{dc} is defined in eq \eq{thermo}. 
We obtain the expression as (see Appendix 5.)   
\begin{eqnarray}
&& \!\!\!\!\!\!\!\!\!\!\!\!\!\!
\tilde{\Pi}_1^{a} (p, \beta, B, \mu_a)_{\rm Thermo} \nn\\
&& = \eta_a q_a^2 l_a^2 p_0(p_0 + 2 m_a) e^{- p_0(p_0 + 2 m_a)l_a^2} 
\left\{ \theta (p_0) - \theta (-p_0 - 2 m_a) \right\} \nn \\
&& \;\; \times 
\left\{
 \fr{\theta (- p_0 - m_a + \mu_a)}{1 + e^{\beta (- p_0 - m_a + \mu_a)}} 
-\fr{\theta (p_0 + m_a - \mu_a)}{1 + e^{\beta (p_0 + m_a - \mu_a)}} 
 + \theta (p_0 + m_a - \mu_a)\right\} \nn\\
&& \; - \eta_a q_a^2 l_a^2 p_0(p_0 - 2 m_a) e^{- l_a^2 p_0(p_0 - 2 m_a)} 
\left\{ \theta (- p_0) - \theta (p_0 - 2 m_a) \right\}\nn \\
&& \;\; \times
\left\{
\fr{\theta (m_a - \mu_a)}{1 + e^{\beta (m_a - \mu_a)}}
- \fr{\theta (- m_a + \mu_a)}{1 + e^{\beta (- m_a + \mu_a)}}
- \theta (m_a - \mu_a) \right\}
+ {\cal O}(|\vec{p}|).   \label{therm2}
\end{eqnarray}
It is easily seen  from eq. \eq{therm2} that the additional 
term $\tilde{\Pi}^{a\nu=1}_1(0)_{\rm Thermo}$ becomes zero at any 
temperature.

Thus, from eqs. \eq{1n0}, \eq{dc} and \eq{dv} with the condition \eq{cc2}, 
we find that eq. \eq{kk} holds when the fermion massless limit is taken or 
when the chirally symmetric case ($N_f^+ = N_f^- = N$) with massive 
(light) fermions is considered.   
In those cases, the following statement is allowed. 
According to the spontaneous magnetization, 
the magnetic part of the field strength has no any static mass and become 
long-range forces.  
If there is no spontaneous magnetization, 
the magnetic field becomes short range force due to  
the non-vanishing topological mass ($\tilde{\kappa} = \kappa \neq 0$).

\section{Discussion}

\hspace*{18pt}
In Sec. 6 we have shown that the spontaneous magnetization induces the
fact that the static masses of magnetic fields vanish and the magnetic
force becomes a long-range one. 
In spite of the direct breakdown of the Lorentz invariance by temperature  
effects, this relation between the magnetization and massless-ness of
the magnetic fields is quite interesting, though the connection with the 
Nambu-Goldstone theorem has not been clarified yet. 
In addition, eq. \eq{kk} is indeed consistent with the result from the
thermodynamical potential approach.  

Now let us consider the reason why the spontaneous magnetization occurs
even at high temperature.  
This result might look somewhat strange when we remember various
symmetries which are spontaneously broken at low temperature and
restored at high temperature; 
chiral symmetry, gauge symmetry for weak interaction, some physics 
of magnetization like in ferromagnetic materials \cite{fem} and so on. 
At $T=0$ case, the viewpoint from observing the shift in zero-point
energies of  photons could explain the appearance of the negative
coefficient of the linear $|B|$ term to some extent \cite{Ho2}.
Similar approximate estimation at finite temperature can be also useful.
In the finite temperature system, the free energies of photons should be 
considered as well as the zero-point energies.
If there is no magnetic fields, the photons have the dynamic mass 
$\sim \sqrt{2N \omega_p^2 + \kappa^2}$ \cite{ft1}.  
On the other hand, if the spontaneous magnetization occurs, the mass 
is shifted to $\sim \sqrt{2N \omega_p^2}$.
The difference of the free energies of the photons between these cases 
is expressed as     
\begin{eqnarray}
  \Delta F^{\rm boson} &=& 
F^{\rm boson}(m^2 = 2N\omega_p^2) - 
F^{\rm boson}(m^2 = 2N\omega_p^2 + \kappa^2)
 \nn \\
&=& \Delta E_{\rm z.p.} + \Delta \tilde{F}^{\rm boson},
\end{eqnarray}
where $\Delta E_{\rm z.p.}$ is deference of the photon zero-point energies and
\begin{eqnarray}
\Delta \tilde{F}^{\rm boson} = 
\fr{1}{\beta} \int \fr{d^2 \vec{p}\;^2}{(2\pi)^2} \ln 
        \fr{1 - e^{-\beta \sqrt{\vec{p}\,^2 + 2N \omega_p^2}}}
           {1 - e^{-\beta \sqrt{\vec{p}\,^2 + 2N \omega_p^2 + \kappa^2}}}. 
\label{dfe}  \end{eqnarray}
The rough estimation for  $\Delta \tilde{F}^{\rm boson}$ shows that 
there is the linear 
$|B|$ term with negative coefficients for small $|B|$ region, 
whereas the linear term in $\Delta E_{\rm z.p.}$ vanishes at high temperature 
limit.
On the other hand,  
it can be seen that the positive contribution of $B$ fields to 
$\Omega_0^{\rm fermi}$
amounts to the order ${\cal O}(|B|^2)$ in the fermion massless limit. 
(When fermions are massive, there is the linear term with positive coefficient 
  at $T = 0$. \cite{Ho3,mir})
These estimations are summarized in Appendix 1. 
Therefore we can realize to some extent that the spontaneous magnetization 
occurs even at finite temperature at least at small $B$ cases.   
  
We have seen that the coefficient $C(\beta)$ 
becomes larger in the high temperature region (see fig. 3). 
We stress that this does not always mean that the spontaneous magnetic field, 
which is determined by observing the minimum of the potential,    
 is an increasing function of $T$. 
Rather, it may be a decreasing function 
if the contribution of the higher orders of $B$ 
to $\Omega^{\rm rad}$ becomes dominant and it holds   
the value of the potential upper as temperature grows in the large $B$ region. 
Then expectation value $\langle B \rangle$ would become smaller, 
even though $C(\beta)$ become larger as $T$ grows.
In this case, $\langle B \rangle$ 
goes to vanish effectively at high temperature limit. 
Such  situation  is very similar to the restoration of the parity violation 
analyzed by Klein-Kreisler {\it et.al.} \cite{KK}.  
We, however, do not discuss any more 
the contribution of higher order of $B$ to the thermodynamical potential 
$\Omega (B,T)$ because our calculation here is basically due to 
the weak field approximation.

When we consider this model in a realistic system such as 
a  planar system in 3+1 dimensional world, 
the spontaneous magnetic field keeps to take non-zero values 
up to a certain critical temperature where a kind of 
the hopping to 3+1 dimension would occur 
and the system would be no longer regarded as a planar system.  
These characteristics may be interesting 
in the connection to the physics such as  the high $T_C$ 
superconductivity \cite{htc}.

\section{Conclusion}

\hspace*{18pt}
In this paper, we have investigated the spontaneous magnetization 
in $QED_3$ with a bare C-S term in the finite temperature system.  
At first, we have analyzed the thermodynamical potential $\Omega$ 
as a function of temperature, chemical potential and magnetic field
in the fermion one-loop level and the massless limit. 
Through these analyses, the consistency condition at finite temperature 
have been assumed, which gives the relation between $\mu$, $|B|$ and $\beta$. 
We have found that the negative coefficient of the linear $B$ term 
is induced in $\Omega$ by radiative correction. 
It means that the spontaneous magnetization can occur even at high 
temperature as well as at zero temperature.  
Secondly the spectrum of the physical photons has been investigated. 
At $T=0$, the photons become massless by radiative effects so that 
they are regarded as the Nambu-Goldstone bosons for the spontaneous 
Lorentz symmetry breaking due to $B \neq 0$.
We found that in our model, the topological mass vanish in the loop level 
even at finite temperature if the classical magnetic field exists. 
Since the magnetic force is a long-range one even at finite temperature unless  
there is a non-vanishing topological mass $\tilde{\Pi}_{\rm topo}$, 
we find that the following statement is allowed.  
According to the spontaneous magnetization, 
the magnetic part of the field strength has no any static mass and become 
long-range forces.  
This statement is consistent with the results by the thermodynamical potential 
approach above. 
Therefore we conclude that the spontaneous magnetic field, 
which is generated at $T=0$,  
does not vanish even at finite temperature. 
This result is valid for the case of the fermion massless limit.

As to the case with massive fermions, we have found that  
$\tilde{\Pi}_{\rm topo}$ vanishes in the chirally symmetric case 
($N^+ = N^- = N$) putting $q_a = e > 0$ and $m_a = m \neq 0$. 
However even in this case it remains unknown whether the spontaneous 
magnetization occurs at finite temperature in the model with massive fermions.  
We have not studied the details of the massive fermion cases  
by thermodynamical potential approach yet. This is a future problem.
These cases would be more important in connection with the fractional 
quantum Hall effects and high $T_C$-superconductivity.

\vspace{1cm}
\noindent
{\large \em Acknowledgments}

Authors would like to thank Profs. K. Higashijima and H. Suzuki 
for useful suggestions and discussions in the early stage of this work,   
Profs. K. Kikkawa, Y. Okada and S. Iso for valuable discussions. 
Thanks are also due to Prof. Y. Hosotani for effective discussions and comments.

\newpage

\appendix
\renewcommand{\theequation}{\thesection.\arabic{equation}}
\setcounter{section}{1}
\addcontentsline{toc}{section}{APPENDICES}
\begin{center}
{\large \bf APPENDICES}
\end{center}
\setcounter{equation}{0}
\subsection*{APPENDIX 1:}

\hspace*{18pt}The Hamiltonian operator, 
$\hat{H}  =  \sum_a \hat{H}_a$, 
and the charge operator, 
$\hat{Q}  =  \sum_a \hat{Q}_a$,
 are given by
\begin{eqnarray}
\hat{H}_a &=& \sum_{n=0}^{\infty} \sum_p a_{np}^{\dagger}a_{np} \omega_n^a
            +\sum_{n=1}^{\infty} \sum_p b_{np}^{\dagger}b_{np} \omega_n^a 
            -\sum_{n=1}^{\infty} \sum_p \omega_n^a , \label{ham}\\
\hat{Q}_a &=& \left\{
\begin{array}{c} q_a \sum_p (a_{0p}^{\dagger}a_{0p}-\frac{1}{2} )\\
                 q_a \sum_p (-b_{0p}^{\dagger}b_{0p}+\frac{1}{2})
\end{array}  \right\}   \nonumber \\
           && \;\;\;\;\;\;\; +
            q_a \sum_{n=1}^{\infty} \sum_p 
            (a_{np}^{\dagger}a_{np} - b_{np}^{\dagger}b_{np})\,,\;\;
           \left\{ \begin{array}{c}  (\eta_a \epsilon (B) > 0)   \\
                                  (\eta_a \epsilon (B) < 0) 
                     \end{array} \right\}. \label{cha}
\end{eqnarray}

The fermion sector of the thermodynamical potential (eq. \eq{fct}) 
is expressed in the classical level as 
\begin{eqnarray}
  \Omega_0^{\rm fermi} &=& 
   \fr{1}{2^{5/2}\pi^2} \zeta \left( \fr{3}{2}\right) \sum_a |q^a B|^{3/2} 
 - \sum_a \fr{|q^a B|}{2\pi \beta} 
   \left\{ \sum_{n=1}^{\infty} 
    \ln \left( 1 + e^{- \beta (\omega_n - q^a\mu^a)} \right) \right. \nn\\
&&  \left. +  \sum_{n=1}^{\infty} 
    \ln \left( 1 + e^{- \beta (\omega_n + q^a\mu^a)} \right)  
  + \ln \left( e^{ \beta \mu_a q_a/2} + 
               e^{- \beta \mu_a q_a/2} \right)\right\},
\label{ccc} 
\end{eqnarray}
where we assumed the fermion massless limit ($m_a = 0$).
In the low temperature limit ($\beta \rightarrow \infty$), 
the second term of r.h.s. vanish and the result in ref. \cite{Ho} 
is reproduced. 
In the case of $\beta^2 |q^a B| \ll 1$, $\Omega_0^{\rm fermi}$
 is calculated as \cite{vol}
\begin{eqnarray}
  \Omega_0^{\rm fermi} &=& \sum_a \left[
     -  \fr{3}{4 \pi \beta^3} \zeta (3)  
+ {\cal O} \left( \beta |q_a B|^2 \right)\right. \nn \\
&& \left. +  \fr{q_a^2}{4\pi} \left\{ \int_0^{\infty} d x 
                \fr{1}{1 + e^{\sqrt{x}}} + {\cal O}(|B|)\right\} \mu_a^2 
   + {\cal O}(\mu_a^4) \right].  \label{ddd}
\end{eqnarray}
We see that there is no linear $B$ term in $\Omega_0^{\rm fermi}$ 
because of eq. \eq{bzm}.

The estimation of $\Delta \tilde{F}$ is approximately performed as follows. 
Since the $\tilde{\Pi}_{1Thermo}$ rapidly becomes 
small for large $|\vec{p}|$ by the factor $e^{- \vec{p}\,^2 l_a^2}$, 
we can employ the cut-off $l_{ave}^{-1}$ for the $|\vec{p}|$ integral in 
$\Delta \tilde{F}^{\rm boson}$, where  
$l_{ave}^{-2} = \sum|\eta_aq_a^3\nu_a|/\sum|\eta_aq_a^2\nu_a| \cdot |B|$ 
is defined in ref. \cite{Ho2}. 
In the case of small $l_{ave}^{-2}/\kappa^2$ (namely small $|B|$), 
we evaluate $\Delta \tilde{F}^{\rm boson}$ as 
\begin{eqnarray}
 \Delta \tilde{F}^{\rm boson}
 &=& \fr{1}{4\pi \beta} \int_0^{l_{ave}^{-2}} d y 
   \ln \fr{1 - e^{- \beta \sqrt{y + 2N \omega_p^2}}}
          {1 - e^{- \beta \sqrt{y + 2N \omega_p^2 + \kappa^2}}} \nn\\ 
&=&  \fr{1}{4\pi \beta} \int_0^{l_{ave}^{-2}} d y 
      \left\{ \fr{1}{2} 
         \ln \fr{y + 2N \omega_p^2}{y + 2N \omega_p^2 + \kappa^2} 
      + {\cal O}(y) \right\}   \nn \\
&=&  - \fr{1}{8\pi \beta} \fr{\kappa^2}{2N \omega_p^2} l_{ave}^{-2}  
                         + {\cal O}( l_{ave}^{-4} ) \nn \\
&=&  - \fr{\kappa e}{8 \pi \ln2}|B| + {\cal O}(|B|^2).
\end{eqnarray}
Thus we can see in this rough estimation 
that the linear $|B|$ term with a negative coefficient appears.

\subsection*{APPENDIX 2:}
\hspace*{18pt}
The fermion propagator in the magnetic field,  
$S(x,y)=-i\langle 0 | {\rm T} \psi (x) \bar{\psi} (y) | 0 \rangle$,
satisfies the relation
\begin{eqnarray}
S(x,y)_{qB<0}= U_C S(y,x)^t_{qB>0}U_C^{\dagger}. \label{s1}  
\end{eqnarray}
The Fourier transformation is given by 
\begin{eqnarray}
S_0(p) = \int \fr{d^3 x}{(2\pi)^3} e^{i p x} S_0(x-y),  \label{sft} 
\end{eqnarray}
where
\begin{eqnarray}
S(x,y)_{qB>0} = e^{-i(x_1-y_1)(x_2+y_2)/2l^2} S_0(x-y)_{qB>0}.
\label{sph}
\end{eqnarray}
The relation of the propagator between $\eta_a=1$ fermions and 
$\eta_a=-1$ ones is     
\begin{eqnarray}
S_0(p)_{\eta_a=-1} = S_0(-p_0, \vec{p})_{\eta_a=+1}.
\label{s2}  \end{eqnarray}
The expression by the proper-time method is convenient in the calculation 
\cite{Ho}. 
For the $\eta_a=1$-fermions, the expression for 
$\nu=0$ is
\begin{eqnarray}
S_0(p)^{\eta=+1}_{\nu=0} &=& 
- i l^2 \int_0^{\infty}d \tau (\cos \tau)^{-1} 
      e^{i(p_0^2 - m^2 + i \epsilon) l^2 \tau - i \vec{p}^{\;2} l^2 \tan \tau}
\nn \\
&& \times  \left\{ \left( m \cos \tau + i p_0 \sin \tau \right) I 
      +  \left(  - i p^1 \sigma_1  - i p^2 \sigma_2 \right) (\cos \tau)^{-1} 
\right.  \nn \\
&& \;\;\;\;\;\;\;\;\;\;\;\;\;\;\;\;\;\;\;\;\;\;\;\;\;\;\;\;\;\;\;\;\;
       \left.     + \left( p_0 \cos \tau + i m \sin \tau \right) \sigma_3
             \right\}. \label{spt}
\end{eqnarray}
As to the case of $\nu=1$, it takes the form 
\begin{eqnarray}
S_0(p)^{\eta=+1}_{\nu=1} &=&  S_0(p)^{\eta=+1}_{\nu=0} + f(p),  
 \label{sdc}
\end{eqnarray}
where $f(p)$ is defined by 
$f(p) \equiv 2 \pi i e^{- \vec{p}^{\;2}l^2} \delta (p_0 - m) (I + \sigma_3)$.

\subsection*{APPENDIX 3: \\
            Boson self-energies, $\Pi_{i}$'s, in the weak field approximation}
\hspace*{18pt}
In our approximation, $\Pi_{i}$'s are formally expressed as 
\begin{eqnarray}
\Pi_0^{a \nu=0}(p, B) 
         &=& \frac{i q_a^2}{\vec{p}^2} 
             \int \frac{d^3 k}{(2\pi)^3}\, \nonumber \\
         &\times&  \frac{ 2\, \left[ m_a^2 + k_0 (k_0 - p_0) + \vec{k}^2 -
                \vec{p} \cdot \vec{k} \right]}
                   {\left( k^2 - m_a^2 + i \epsilon \right)
                    \left[(k - p)^2   - m_a^2 + i \epsilon \right]}+O(B^2), 
\label{n1}   \\
\Pi_1^{a \nu=0}(p, B) 
         &=& \frac{2 i \eta_a q_a^2}{\vec{p}^2} 
              \int \frac{d^3 k}{(2\pi)^3}\, \nonumber \\
         &\times&  \frac{  m_a \vec{p}^2 - i (p_0 - 2 k_0)
                             (k_1p_2 -k_2 p_1) }
                   {\left( k^2 - m_a^2 + i \epsilon \right)
                    \left[(k - p)^2 - m_a^2 + i \epsilon \right]}+O(B^2), 
\label{n2}   \\
\Pi_2^{a \nu=0}(p, B) 
  &=& \;  \frac{4 i q_a^2}{(\vec{p}^2)^2}
                \int \frac{d^3 k}{(2\pi)^3} \nonumber \\
  & & \times \frac{p_0^2 \left[ m_a^2  + k_0(k_0-p_0) + \vec{k}^2 
                           - \vec{p} \cdot \vec{k} \right]  
                + \vec{p}^2 \left[ m_a^2 - k_0(k_0 - p_0) \right]}
             {\left( k^2 - m_a^2  + i \epsilon \right)
              \left[ (k - p)^2  - m_a^2 + i \epsilon    \right]} \nn \\
&&  \;\;\;\;\;\;\;\;\;\;\;\;\;\;\;\;\;\;\;\;\;\;\;\;\;\;\;\;\;\;
   \;\;\;\;\;\;\;\;\;\;\;\;\;\;\;\;\;\;\;\;\;\;\;\;\;\;\;\;\;\;\;\;\;
 + O(B^2),  
\label{n3}\\
\Pi_0^{a \nu=1}(p, B) 
       &=&  \Pi_0^{a \nu=0}(p, B)\, \nonumber \\
       & &\!\!\!\!\!\!\!\!\!\!\!\!
     -  \frac{4\pi q_a^2}{\vec{p}^2}
        \int \fr{d^3 k}{(2\pi)^3} 
\fr{m_a + k_0}{k_0^2 - \vec{k}\,^2 - m_a^2 + i \epsilon}     \nn\\
&& \!\!\!\!\!\!\!\!\!\!\!\! \times 
  \left\{   e^{-(\vec{k} - \vec{p})^2l_a^2}\, \delta (k_0 - p_0 - m) 
 +e^{- (\vec{k} + \vec{p})^2l_a^2}\,\delta (k_0 + p_0 - m)\right\},
\label{n4}\\
\Pi_1^{a \nu=1}(p, B) 
      &=& \; \Pi_1^{\nu=0}(p, B)
 \nonumber     \\
      & &  \!\!\!\!\!\!  + \frac{4 \pi \eta_a q_a^2}{\vec{p}^2}
   \int \frac{d^3 k}{(2\pi)^3}\,
            \nonumber    \\  
& & \!\!\!\!\!\! \times \left[ \frac{(ip_1 - p_2)k_2 - (p_1 + i p_2)k_1}
          {k_0^2 - \vec{k}^2 -m_a^2 + i \epsilon}  \,
     \delta (k_0 - p_0 - m_a) e^{-(\vec{k} - \vec{p})^2 l_a^2}\right. 
           \nonumber \\
& &  \!\!\!\!\!\!\!\!\!\!
    \left. + \frac{-\vec{p}^2+(p_1 - i p_2)k_1+(i p_1 + p_2)k_2}
            {(k - p)^2 -m_a^2 + i\epsilon} \,
      \delta (k_0 - m_a) e^{- \vec{k}^2 l_a^2}\right], \label{n5} \\
  \Pi_2^{a \nu=1}(p, B)  &=& 
  \Pi_2^{a \nu=0}(p, B)\, \label{n6}.
\end{eqnarray}
The straightforward calculation of \eq{n1} $\sim$ \eq{n6} reproduces 
the  results in \cite{ch2,Ho} up to the order ${\cal O}(B_J^1)$; 
\begin{eqnarray}
  \Pi_0^{\nu=0}  &=&  \fr{q_a^2}{8\pi} \fr{1}{(- p^2)^{1/2}} 
\left\{ \sqrt{z} + (1 - z) \sin^{-1} \fr{1}{\sqrt{1+z}}\right\},   \\
  \Pi_1^{\nu=0} 
&=& - \fr{\eta_a q_a^2}{4\pi} \sqrt{z} \sin^{-1} \fr{1}{\sqrt{1 + z}}, \\
 \Pi_2^{\nu=0} 
&=& \fr{q_a^2}{8\pi} \fr{1}{(- p^2)^{1/2}} 
         \left( \sqrt{z} + (1 - z) \sin^{-1} \fr{1}{\sqrt{1 + z}}\right),  
\end{eqnarray}
where $z = - 4m_a^2/p^2$ and 
\begin{eqnarray}
\Pi^{a \nu=1}_0  
&=& \Pi^{a \nu=0}_0  - \fr{q_a^2}{2\pi} \fr{1}{p_0 l_a^2} 
\left\{ \fr{1}{p^2 + 2 m_a p_0} - \fr{1}{p^2 - 2 m_a p_0}\right\},  \\ 
\Pi^{a \nu=1}_1 
&=&  \Pi^{a \nu=0}_1- \fr{q_a^2}{2\pi} \fr{\eta_a}{l_a^2} 
\left\{ \fr{1}{p^2 + 2 m_a p_0} + \fr{1}{p^2 - 2 m_a p_0}\right\}, \label{qq1} \\
\Pi^{a \nu=1}_2  
&=&  \Pi^{a \nu=0}_2.
\end{eqnarray}

\subsection*{APPENDIX 4:\\ 
 $\tilde{\Pi}_0$, $\tilde{\Pi}_1$ and $\tilde{\Pi}_2$ 
 in weak field approximation}
\hspace*{18pt}
By using the replacement \eq{mm3} and \eq{mm4}, the self-energies in the 
finite temperature system are formally obtained 
from eqs. \eq{n1} $\sim$ \eq{n6}. 
After performing the angular integral included in $d \vec{k}$, 
the expression of $\tilde{\Pi}_i$'s takes the following form.  
For the $\nu = 0$ case,
\begin{eqnarray}
&&\tilde{\Pi}_0^{\nu=0}(p_3, \vec{p}^2, B, \beta)  
= - \fr{q^2}{\vec{p}^2}\fr{1}{2\pi \beta} \sum_{n=-\infty}^{\infty} 
  \int_0^{\infty} d k \fr{k}{k_3^2 + k^2 + m^2}  \nn \\
&& \;\;\;\; \times 
    \left\{ 1 + \fr{k^2 + m^2 - p^2 - (k_3 - p_3)(3 k_3 - p_3)}
         {\sqrt{
          \left\{(k + |\vec{p}|)^2 + m^2 + (k_3 - p_3)^2 \right\}
    \left\{(k - |\vec{p}|)^2 + m^2 + (k_3 - p_3)^2 \right\}}}\right\}  \nn\\
&& \;\;\;\;  + \;{\cal O}(B^2), \\ 
&&\tilde{\Pi}_1^{\nu=0}(p_3, \vec{p}^2, B, \beta) = 
\fr{-  q^2 \eta}{2\pi} \fr{m}{\beta} 
 \sum_{n= -\infty}^{\infty} \int_0^1 d x \fr{1}{m^2 + (k_3 - p_3 x)^2 
                                                - x (1 - x) \vec{p}^2} \nn \\
&& \;\;\;\; + \; {\cal O}(B^2), \label{nu0} \\
&& \tilde{\Pi}_2^{\nu=0}(p_3, \vec{p}^2, B, \beta) \nn \\
&&\;\;=  - \fr{p_3^2}{\vec{p}^2} \tilde{\Pi}_0^{\nu=0}
(p_3, \vec{p}^2, B, \beta)_m 
 + \fr{q^2}{\vec{p}^4}\fr{1}{2 \pi \beta} \sum_{n=- \infty}^{\infty}
   \int_0^{\infty} d k \fr{k}{k_3^2 + k^2}   \nn \\
&& \;\;\;\; \times \left[ p_3^2  + \fr{p_3^2  k^2 - 
     (4 p^2 + 3 p_3^2)k_3^2 + (4 k_3 - p_3)(p^2 + p_3^2)p_3
              - (4 \vec{p}^2 - p_3^2) m^2}
     {\sqrt{          \left\{(k + |\vec{p}|)^2 + m^2 + (k_3 - p_3)^2 \right\}
    \left\{(k - |\vec{p}|)^2 + m^2 + (k_3 - p_3)^2 \right\}}}\right] \nn \\
&& \;\;\;\; + \;{\cal O}(B^2).
\end{eqnarray}
As to the case of $\nu =1$,
\begin{eqnarray}
\tilde{\Pi}_0^{\nu=1}(p_3, \vec{p}^2, B, \beta) 
&=&  \tilde{\Pi}_0^{\nu=0}(p_3, \vec{p}^2, B, \beta) \nn \\
+\;\;\;\;\;\;\;\;\;\;\;\;&& \!\!\!\!\!\!\!\!\!\!\!\!\!\!\!\!\!\!\!\!\!\!\!\!
\!\!\!\!\!\!\!\!\! \fr{q^2}{2 \pi l^2  p_3}
\left\{   \fr{1}{p_3^2 + \vec{p}^2 - i 2 p_3 m}  
        - \fr{1}{p_3^2 + \vec{p}^2 + i 2 p_3 m} \right\}
 + m \times{\cal O}(e^{- \vec{p}\,^2/(q_a|B|)}),\nn \\
&& \\
\tilde{\Pi}_1^{\nu=1}(p_3, \vec{p}^2, B, \beta) 
&=&  \tilde{\Pi}_1^{\nu=0}(p_3, \vec{p}^2, B, \beta) \nn \\
+\;\;\;\;\;\;\;\;\;\;\;\;&& \!\!\!\!\!\!\!\!\!\!\!\!\!\!\!\!\!\!\!\!\!\!\!\!
\!\!\!\!\!\!\!\!\!
 \fr{q^2 \eta}{2 \pi l^2}
\left\{   \fr{1}{p_3^2 + \vec{p}^2 - i 2 p_3 m}  
        + \fr{1}{p_3^2 + \vec{p}^2 + i 2 p_3 m} \right\} 
 + {\cal O}(e^{- \vec{p}\,^2/(q_a|B|)}),  \nn \\
&& \\
\tilde{\Pi}_2^{\nu=1}(p_3, \vec{p}^2, B, \beta) 
&=&  \tilde{\Pi}_2^{\nu=0}(p_3, \vec{p}^2, B, \beta).
\end{eqnarray}

\subsection*{APPENDIX 5:  Calculation of $\tilde{\Pi}_{1\rm Thermo}(p = 0)$}
\hspace*{18pt}The second term in r.h.s. in eq. \eq{dc} is given  by 
\begin{eqnarray}
\tilde{\Pi}_{1 {\rm Thermo}} (p, B, \beta, \mu) \nn 
&=& \fr{-1}{2\pi i} 
\fr{2 \eta_a q_a^2}{\vec{p}^2} \int_{- \infty}^{\infty} d \lambda 
\int \fr{d^2 \vec{k}}{(2\pi)^2} \nn \\
&& \!\!\!\!\!\!\!\!\!\!\!\!\times
\left\{ \int_{-i\infty + \mu + \epsilon}^{i \infty + \mu + \epsilon}
         d k_0 f(k_0, \lambda)  \fr{1}{e^{\beta (k_0 - \mu)} + 1} \right. \nn \\
&& \!\!\!\!\!\!\!\!\!\!\!\!
 + \left.  \int_{-i\infty + \mu - \epsilon}^{i \infty + \mu - \epsilon}  
         d k_0 f(k_0, \lambda)  \fr{1}{e^{- \beta (k_0 - \mu)} + 1} 
       - \oint_C  d k_0 f(k_0, \lambda)    \right\},
\label{thermo} \nn\\ 
\end{eqnarray}
where 
\begin{eqnarray}
  f(k_0, \lambda) 
&\equiv& \fr{(i p_1 - p_2)k_2 - (p_1 + i p_2)k_1}
            {k_0^2 - \vec{k}^2 - m^2 + i \epsilon}
e^{i (k_0 - p_0 - m) \lambda} e^{- (\vec{k} - \vec{p})^2 l_a^2} \nn \\
&& \fr{- \vec{p}^2 + (p_1 - i p_2) k_1 + (i p_1 + p_2)k_2}
           {(k_0 - p_0)^2 - (\vec{k} - \vec{p})^2 - m^2 + i \epsilon}
e^{i (k_0 -  m) \lambda} e^{- \vec{k}^2 l_a^2}.  
\end{eqnarray}
The straightforward calculation produces   
\begin{eqnarray}
&& \!\!\!\!\! \tilde{\Pi}_{1{\rm Thermo}} (p, B, \beta, \mu) \nn\\
&&= +\fr{\eta_a q_a^2}{2\pi} \fr{p_1 +  i p_2}{\vec{p}\,^2}
    \sqrt{p_0 (p_0 + 2 m_a)} 
    \left\{ \theta (p_0) - \theta (-p_0 - 2 m_a) \right\} \nn \\
&& \times \left\{
 \fr{\theta (- p_0 - m_a + \mu_a)}{1 + e^{\beta (- p_0 - m_a + \mu_a)}} 
-\fr{\theta (p_0 + m_a - \mu_a)}{1 + e^{\beta (p_0 + m_a - \mu_a)}} 
 + \theta (p_0 + m_a - \mu_a)\right\}\nn\\
&& \times 
    e^{-\vec{p}\,^2 l^2}e^{- p_0(p_0 + 2m_a) l^2} 
\int_0^{2\pi} d \phi  e^{- i \phi} 
        e^{2 l^2 \sqrt{p_0(p_0 + 2m_a)} 
(p_1 \cos \phi + p_2 \sin \phi)}\nn \\
&& + \fr{\eta_a q_a^2}{2\pi} \fr{p_1 -  i p_2}{\vec{p}^2}
    \sqrt{p_0 (p_0 - 2 m_a)}
\left\{ \theta (- p_0) - \theta (p_0 - 2 m_a) \right\}  \nn \\
&& \times \left\{
\fr{\theta (m_a - \mu_a)}{1 + e^{\beta (m_a - \mu_a)}}
- \fr{\theta (- m_a + \mu_a)}{1 + e^{\beta (- m_a + \mu_a)}}
- \theta (m_a - \mu_a) \right\}\nn\\
&& \times  e^{-\vec{p}^2 l^2}e^{- p_0(p_0 - 2m_a) l^2}
\int_0^{2\pi} d \phi  e^{ i \phi} 
        e^{-2 l^2 \sqrt{p_0(p_0 - 2m_a)} (p_1 \cos \phi + p_2 \sin \phi)}
\label{therm}  \end{eqnarray}
To perform the integral for $d \phi$ in eq. \eq{therm}, 
we expand the integrants by $|\vec{p}|$.  Then the integrations 
become 
\begin{eqnarray}
&&  \int_0^{2\pi} d \phi  e^{ \mp i \phi} 
        e^{\pm2 l^2 \sqrt{p_0(p_0 \pm 2m)} (p_1 \cos \phi + p_2 \sin \phi)}.
\nn \\
&& \;\;\;\;\;\;\;\;\;\;\;\;\;\;\;\; 
 = \pm 2\pi l^2 \sqrt{p_0(p_0 \pm 2m)} (p_1 \mp i p_2) + {\cal O}(\vec{p}\;^2). 
\label{ff}
\end{eqnarray}
By using eq. \eq{ff}, we finally obtain eq. \eq{therm2} from eq. \eq{therm}.

\subsection*{APPENDIX 6}
\hspace*{18pt}
The formula to divide functions into the zero-temperature part and additional 
finite temperature part is \cite{ft1} 
\begin{figure}[b]
\centering{
\leavevmode
\psfig{file=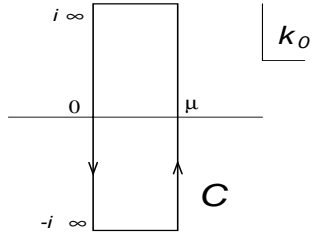,height=40mm,width=60mm,angle=-90}
}
\caption{The contour $C$ for the integral in eq. (A.34).}
\end{figure}
\begin{eqnarray}
\fr{1}{\beta} \sum_n f (k_0) &=&  \fr{1}{2\pi i}
     \int_{-i \infty}^{i \infty} d k_0 f(k_0) 
+ \fr{1}{2\pi i} \oint_C d k_0 f(k_0)\nn \\
 &&  - \fr{1}{2\pi i}
    \int_{-i \infty + \mu + \epsilon}^{i \infty + \mu +\epsilon} 
       \fr{d k_0 f(k_0)}{e^{\beta (k_0 - \mu)} + 1} - \fr{1}{2\pi i}
    \int_{-i \infty + \mu - \epsilon}^{i \infty + \mu - \epsilon} 
       \fr{d k_0 f(k_0)}{e^{- \beta (k_0 - \mu)} + 1}, \nn \\
\label{def}
\end{eqnarray}
where $f(k_0)$ is a function without having 
singularities on imaginary axis and 
contour $C$ is defined in Fig. 4.

Euler-Mclaurin's formula is given by  
\begin{eqnarray}
&&\sum_{k=0}^n g(a + h k) =  
\fr{1}{h} \int_a^{a+nh} g(x) dx 
           + \fr{1}{2} \left\{ g(a) +  g(a+nh) \right\} \nn \\
 &&  \;\;
            + \sum_{r=1}^{m-1}(-1)^{r-1} \fr{Br}{(2 r)!} h^{2r-1} 
                \left\{ g^{(2r-1)}(a + n h) -  g^{(2r-1)}(a) \right\} 
                                             + {\cal O}(h^{2m}),\nn \\
\end{eqnarray}
where $B_r \equiv 2(2r)!/(2^{2r}-1) \pi^{-2r} \sum_{n=1}^{\infty}(2n+1)^{-2r}$  
(the Bernoulli numbers).

\newpage


\begin{thebibliography}{99}
\bibitem{qh1} K. Ishikawa, Phys. Rev. D 31,  1432 (1985); 
              N. Imai, K. Ishikawa, T. Matsuyama, and I. Tanaka, 
                          Phys.Rev. B 42, 10610 (1990); 
              V. Zeitlin, Phys. Lett. B 352, 422 (1995). 
\bibitem{qh2} J. K. Jain, Phys. Rev. Lett. 63, 199 (1989); 
              T. Chakraborty and P. Pietil\"{a}inen, 
                   {\it The Fractional Quantum Hall Effect},
                   (Springer-Verlag, 1988)
\bibitem{htc} R. B. Laughlin, Phys. Rev. Lett. 60, 2677 (1988);
              J. D. Lykken, J. Sonnenschein, and N. Weiss, 
                      Int. J. Mod. Phys. A 6, 5155 (1991).
\bibitem{ch1} M. R. Pennington and D. Walsh, Phys. Lett. B 253, 246 (1991);
              M. C. Diamantini, P. Sodano, and G. W. Semenoff, 
                      Phys. Rev. Lett. 70, 3848 (1993);  
              D.Nash, Phys. Rev. Lett. 62, 3024 (1989). 
\bibitem{ch2}
              R. D. Pisarski, Phys. Rev. D 29, 2423 (1984);
              T. W. Appelquist, M. J. Bowick, D. Karabali, 
                                    and L. C. R. Wijewardhana, 
                     Phys. Rev. D 33, 3704 (1986).
\bibitem{pb}  A. N. Redlich, Phys. Rev. D 29, 2366 (1984); 
              T. W. Appelquist, M. J. Bowick, D. Karabali, 
                                    and L. C. R. Wijewardhana, 
                     Phys. Rev. D 33, 3774, {\it brief reports} (1986). 
\bibitem{Ho}    Y. Hosotani, Phys. Lett. B 319, 332 (1993).
\bibitem{Ho2}   Y. Hosotani, Phys. Rev. D 51, 2022 (1995).     
\bibitem{IS}    T. Itoh and T. Sato, Phys. Lett. B 367, 290 (1996).
\bibitem{Ho3}   D. Wesolowski and Y. Hosotani,
                       Phys. Lett. B 354, 396 (1995);\\
                V. Zeitlin, 
                       Preprint FIAN/TD/96-09, ({\tt hep-th/9605194}). 
\bibitem{br} N. Brali\'{c}, D. Cabra, and F. A. Schaposnik, 
                   Phys. Rev. D 50, 5314 (1994).
\bibitem{KK} M. Klein-Kreisler and M. Torres, 
             Phys. Lett. B 347, 361 (1995).
\bibitem{km1} S. Kanemura and T. Matsushita, 
             Preprint OU-HET 212, \\({\tt hep-th/9505146}).
\bibitem{cht} S. Coleman and B. Hill, Phys. Lett. B 159, 184 (1985);
              G. W. Semenoff, P. Sodano, and Y-S. Wu, 
              Phys. Rev. Lett. 62, 715 (1989). 
\bibitem{bal} T. Banks and J. D. Lykken, Nucl. Phys. B 336, 500 (1990).
\bibitem{lin} A. D. Linde, Rep. Prog. Phys., Vol.42, 389 (1979).
\bibitem{ft1} J. I. Kapusta, {\it Finite-Temperature Field Theory}, 
                 (Cambridge University Press, 1989) 
\bibitem{wel} H. A. Weldom, Phys. Rev. D 26, 1394 (1982).
\bibitem{lag} A. I. Akheizer and V. B. Berestetsky, 
              {\it Quantum Electrodynamics}, 
              (Interscience, New York, 1965). 
\bibitem{lan} L. D. Landau and E. M. Lifshiz, 
             {\it Statistical Physics}, (Pergamon Press, 1958).
\bibitem{mm}  T. Matsubara, Prog. Theor. Phys. 14, 351 (1955).
\bibitem{dj}  L. Dolan and R. Jackiw, Phys. Rev. D 9, 3320 (1974).
\bibitem{fem}   
       {\it Magnetic Phase Transitions},  (Springer-Verlag. Editors:  
             M. Ausloos and R. J. Elliott, 1983) 
\bibitem{lsw} J. D. Lykken, J. Sonnenschein, and N. Weiss, 
               Phys. Rev. D 42, 2161 (1990). 
\bibitem{mir} V. P. Gusynin, V. A. Miransky, and I. A. Shovkovy, 
              Phys. Rev. D 52, 4718 (1995). 
\bibitem{vol} M. B. Voloshin, Preprint TPI-MINN-96/14-T, UMN-TH-1508-96,\\ 
                        ({\tt hep-ph 9609219}). 
\end{thebibliography}
\end{document}